\documentclass[aap,MSNbibl,seceqn,citesort,dvips]{arximspdf}

\doi{10.1214/11-AAP825} 
\volume{23}
\issue{2}
\pubyear{2013}
\firstpage{427}
\lastpage{454}

\makeatletter

\newcommand{\cal}{\mathcal}
\newcommand{\textbff}{}

\newtheorem{Theorem}{Theorem}[section]
\newproclaim{Definition}{Definition}[section]
\newtheorem{Proposition}{Proposition}[section]
\newproclaim{Assumption}{Assumption}[section]
\newtheorem{Lemma}{Lemma}[section]
\newtheorem{Corollary}{Corollary}[section]
\newproclaim{Remark}{Remark}[section]

\newproclaim{Condition}{Condition}[section]

\newcommand{\esssup}{\mathop{\operatorname{ess}\operatorname{sup}}}

\newcommand{\sesq}[2]{\langle #1, #2 \rangle}

\newcommand{\E}{\mathbb{E}}
\newcommand{\F}{\mathbb{F}}
\newcommand{\M}{\mathbb{M}}
\newcommand{\N}{\mathbb{N}}
\newcommand{\R}{\mathbb{R}}

\renewcommand{\P}{\mathbb{P}}
\newcommand{\Q}{\mathbb{Q}}

\newcommand{\T}{\mathbb{T}}
\newcommand{\Ac}{{\cal A}}
\newcommand{\Bc}{{\cal B}}

\newcommand{\Ec}{{\cal E}}
\newcommand{\Fc}{{\cal F}}
\newcommand{\Gc}{{\cal G}}

\newcommand{\Ic}{{\cal I}}

\newcommand{\Mc}{{\cal M}}
\newcommand{\Xc}{{\cal X}}
\newcommand{\Yc}{{\cal Y}}

\newcommand{\eps}{\varepsilon}

\newcommand{\one}{{\mathbf1}}

\makeatother

\begin{document}
\begin{frontmatter}

\title{No-arbitrage of second kind in countable markets with proportional transaction costs}
\runtitle{No-arbitrage of second kind in countable markets}

\begin{aug}
\author[A]{\fnms{Bruno} \snm{Bouchard}\ead[label=e1]{bouchard@ceremade.dauphine.fr}}
\and
\author[B]{\fnms{Erik} \snm{Taflin}\corref{}\ead[label=e2]{taflin@eisti.fr}}
\runauthor{B. Bouchard and E. Taflin}
\affiliation{Universit\'e Paris Dauphine and EISTI}
\address[A]{CEREMADE\\
Universit\'e Paris Dauphine\\
CREST, ENSAE\\
France \\
\printead{e1}}
\address[B]{EISTI \\
Avenue du Parc \\
95011 Cergy \\
France \\
\printead{e2}}
\end{aug}


\received{\smonth{10} \syear{2010}}
\revised{\smonth{10} \syear{2011}}

%
\begin{abstract}
Motivated by applications to bond markets, we propose a multivariate
framework for discrete time financial markets with proportional
transaction costs and a countable infinite number of tradable assets.
We show that the no-arbitrage of second kind property (NA2 in short),
recently introduced by R\'{a}sonyi
for finite-dimensional markets, allows us to provide a closure property for
the set of attainable claims in a very natural way, under a suitable
efficient friction condition. We also extend to this context the
equivalence between NA2 and the existence of many (strictly) consistent
price systems.
\end{abstract}

%

\end{frontmatter}

\section{Introduction}
\label{secintroduction}

Motivated by applications to bonds markets, for which it is
acknowledged that all possible maturities have to be taken into
account, many papers have been devoted to the study of financial models
with infinitely many risky assets; see, for example,
\cite{BMKR97,Carmona-Tehr,DP05b,ET,Pham2003} and the
references therein. To the best of our knowledge, models with
proportional transaction costs have not been discussed so far. This
paper is a first attempt to treat such situations in a general
framework.

As a first step, we restrict to a discrete time setting where a
{countable infinite} number of financial assets is available. Time
belongs to $\T:=\{0,\ldots,T\}$.

Following the modern literature on financial models with proportional
transaction costs (see~\cite{Kab-Saf} for a survey), financial
strategies are described here by $\R^\N$-valued $(\Fc_t)_{t\in\T
}$-adapted processes $\xi=(\xi_t)_{t\in\T}$, where $(\Fc_t)_{t\in
\T}$ is a given filtration that models the flow of available
information, and each component $\xi^i_t$ of $\xi_t=(\xi^i_t)_{i\ge
1} \in\R^\N$ describes the changes in the position on the financial
asset $i$ induced by trading on the market at time $t$.

When the number of financial assets is finite, say $d$, one can view
each component $\xi^i_t$ as the amount of money invested in the asset
$i$ or as a number of units of asset $i$ held in the portfolio.

The main advantage of working in terms of units is that it is num\'
eraire free; see the discussions in~\cite{KSR04} and~\cite{schach}.
In such models, the self-financing condition is described by a cone
valued process $\hat K=(\hat K_t)_{t\in\T}$ which incorporates
bid-ask prices. Namely, a financial strategy is said to satisfy the
self-financing condition if $\xi_t\in-\hat K_t$ a.s. for
all $t\in
\T$, where $-\hat K_t(\omega):=\{y\in\R^d\dvtx y^i\le\sum_{i\neq
j}(a^{ji}-a^{ij}\pi^{ij}_t(\omega)),\forall i\le d$, for some
$a=(a^{ij})_{i,j\ge1}\in\R^{d\times d}$ with nonnegative entries\}.
In the above formulation, $\pi^{ij}_t$ stands for the number of units
of asset $i$ required in order to buy one unit of asset $j$ at time
$t$. The self-financing condition then just means that the changes $\xi
_t$ in the portfolio can be financed (in the large sense) by passing
exchange orders $(a^{ij})_{i,j\ge1}$ on the market, that is,
$a^{ij}\ge0$ represents the number of units of asset $j$ that are
obtained against $a^{ij}\pi^{ij}_t$ units of asset $i$.

Under the so-called \textit{efficient friction} assumption, namely $\pi
^{ij}_t\pi^{ji}_t>1$ for all $i,j$ and $t\le T$, and under suitable no
arbitrage conditions (e.g., the strict no-arbitrage condition of \cite
{KSR01} or the robust no-arbitrage\vspace*{1pt} condition of~\cite{schach}; see
also~\cite{KSR04}), one can show that there exists a martingale $\hat
Z=(\hat Z_t)_{t\le T}$ such that,\vspace*{1pt} for all $t\le T$, $\hat Z_t$ lies in
the interior of the (positive) dual cone $\hat K^*_t$ of $\hat K_t$,
which turns out to be given by
\[
\hat K^*_t(\omega)=\{z\in\R^d\dvtx0\le z^j\le z^i \pi^{ij}_t(\omega
), i,j\le d\}.
\]
The martingale $\hat Z$ has then the usual interpretation of being
associated to a fictitious frictionless market which is cheaper than
the original one, that is, $\hat Z^j_t/\hat Z^i_t<\pi^{ij}_t$, and
such that the classical no-arbitrage condition holds, that is, $\hat Z$
is a martingale. This generalizes to the multivariate setting the
seminal result of~\cite{JKequi}.

The existence of such a martingale can then be extended to the
continuous setting (see~\cite{GuRaSc08} for a direct approach in a
one-dimensional setting and~\cite{GrepaKaba} for a multivariate
extension based on a discrete time approximation), which, in turn,
allows us to prove that the set of attainable claims is closed is some
sense; see, for example, Lemma 12 and the proof of Theorem 15 in \cite
{campischach}; see also~\cite{BoCh09} and~\cite{DeDeKa}. Such a
property is highly desirable when one is interested by the formulation
of a dual representation for the set of super-hedgeable claims, or by
existence results in optimal portfolio management; see the above papers
and the references therein.

The aim of this paper is to propose a generalized version of the above
results to the context of discrete time models with a {countable
infinite} number of assets, with the purpose of providing later a
continuous time version.

When the number of assets is {countable infinite}, the first difficulty
comes from the notion of interior associated to the sequence of dual
cones $(\hat K_t^{*})_{t\in\T}$. Indeed, a~natural choice would be to
define $\hat K_t(\omega)$ as a subset of $l^1$, the set of elements
$x=(x^i)_{i\ge1}\in\R^\N$ such that $|x|_{l^1}:=\sum_{i\ge1}
|x^i|<\infty$, so as to avoid having an infinite global position in a
subset of financial assets; see~\cite{T09} for a related criticism on
frictionless continuous time models. In this case, $\hat K^*_t$ should
be defined in $l^\infty$, the set of elements $x=(x^i)_{i\ge1}\in\R
^\N$ such that $|x|_{l^\infty}:=\sup_{i\ge1} |x^i|<\infty$. But,
for the topology induced by $|\cdot|_{l^\infty}$, the sets $\hat
K^*_s(\omega)$ have no reason to have a nonempty interior, except
under very strong conditions on the bid-ask matrices $(\pi
^{ij}_t(\omega))_{i,j}$.

We therefore come back to the original modelization of~\cite{KSR01} in
which financial strategies are described through amounts of money
invested in the different risky assets. Namely, we assume that the
bid-ask matrix $(\pi^{ij}_t)_{i,j}$ takes the form $((1+\lambda
^{ij}_t)S^j_t/S^i_t)_{i,j}$ where $S^k_t$ stands for the price, in some
num\'eraire, of the risky asset $k$, and $\lambda^{ij}_t$ is a positive
coefficient (typically less than $1$) interpreted as a proportional
transaction cost. The changes $\xi_t$ in the portfolio due to trading
at time~$t$, now quoted in terms of the num\'eraire, thus take values
in the set $-K_t$ where $K_t(\omega):=\{(S^i_t(\omega) y^i)_{i\ge1},
y\in\hat K_t(\omega)\}$. Viewed as a subset of $l^1$, $K_t(\omega)$ has
a dual cone $K^*_t(\omega)\subset l^\infty$ which takes the
form\looseness=-1
\[
K^*_t(\omega):=\bigl\{z\in l^\infty\dvtx0\le z^j\le z^i \bigl(1+\lambda
^{ij}_t(\omega)\bigr), i,j\ge1\bigr\},
\]\looseness=0
and whose interior in $l^\infty$ is now nonempty under mild
assumptions, for example, if $\lambda^{ij}_t(\omega)\ge\eps(\omega
)$ a.s. for all $i,j\ge1$ for some random variable $\eps$ taking
strictly positive values.

This approach, although not num\'eraire free, allows us to bound the
global amount invested in the different subsets of assets, by viewing
$K_t$ as a subset of~$l^1$, while leaving open the possibility of
finding a process $Z$ such that such $Z_t$ lies in the interior of
$K^*_t$ a.s., that is, such that $\hat Z:=ZS$ still satisfies $\hat
Z^j_t/\hat Z^i_t<\pi^{ij}_t$ for all $i,j$.

We shall see below that, under a suitable no-arbitrage condition, one
can actually choose $Z$ in such a way that $ZS$ is a martingale, thus
recovering the above interpretation in terms of arbitrage free
fictitious market. Moreover, we shall show that the set of terminal
wealths induced by financial strategies defined as above is indeed
closed in a suitable sense; see Theorems~\ref{thmSTFatouclosure} and
\ref{thmfermeturefaible}. This means that we do not need to
consider an additional closure operation in order to build a nice
duality theory or to discuss optimal portfolio management problems, as
it is the case in frictionless markets; cf.~\cite{ETBondCompleteness}
and~\cite{T09} for a comparison with continuous time settings.

Another difficulty actually comes from the notion of no-arbitrage to be
used in such a context. First, we should note that various, a priori
not equivalent, notions of no-arbitrage opportunities can be used in
models with proportional transaction costs. We refer to~\cite{Kab-Saf}
for a complete presentation and only mention one important point: the
proofs of the closure properties, of the set of attainable claims,
obtained in~\cite{KSR01} and~\cite{schach}, under the strict
no-arbitrage and the robust no-arbitrage property, heavily rely on the
fact that the boundary of the unit ball is closed in $\R^d$ (for the
pointwise convergence). This is no more true, for the pointwise
convergence, when working in $l^1$ viewed as a subspace of $\R^\N$
with unit ball defined with $|\cdot|_{l^1}$. In particular, it does
not seem that they can be reproduced in our infinite-dimensional setting.

However, we shall show that the notion of no-arbitrage of second kind
(in short NA2), recently introduced by~\cite{ras09} under the label
``no-sure profit in liquidation value,'' is perfectly adapted. It says
that the terminal value $V_T$ of a wealth process cannot take values
a.s. in $K_T$ if the wealth process at time $t$, $V_t$, does not
already take values a.s. in $K_t$, for $t\le T$. Note that $V_t\in K_t$
if and only if $-V_t\in-K_t $. Since $V_t+(-V_t)=0$, this means that
$K_t$ is the set of position holdings at time $t$ that can be turned
into a zero position, after possibly throwing away nonnegative amounts
of financial assets, that is, $K_t$ is the set of ``solvable''
positions at time $t$. Hence, the NA2 condition means that we cannot
end up with a portfolio which is a.s. solvable if this was not the case
before, which is a reasonable condition.

Under this condition, we shall see that a closure property can be
proved under the 
assumption 
that $K^*_t$ has a.s. a nonempty interior, for all $t\le T$, which is,
for instance, the case if $\eps\le\lambda^{ij}_t(\omega)\le\eps
^{-1} $ a.s. for all $i,j\ge1$ and $t\le T$, for some \mbox{$\eps>0$}.
We shall also extend to our framework the PCE (Prices Consistently
Extendable) property introduced in~\cite{ras09}, which we shall call
MSCPS (Many Strictly Consistent Price Systems) to follow the
terminology of~\cite{DeKa10}.

The rest of the paper is organized as follows.
We first conclude this Introduction with a list of notation that will
be used throughout paper. The model and our key assumptions are
presented in Section~\ref{S1}. Our main results are reported in
Section~\ref{secmainresults}. The proofs of the closure properties
are collected in Section~\ref{secclosure}, in which we also prove a
dual characterization for the set of attainable claims and discuss the
so-called B-property. The existence of 
Many Strictly Consistent Price Systems is proved in Section \ref
{secmcps}. We then discuss elementary properties of cones in
infinite-dimensional spaces and under which conditions our key
assumption, Assumption~\ref{EffFric} below, holds. Finally, in Section
\ref{secconcludingremars}, we explain how our results can be
generalized to a more abstract setting.\vspace*{10pt}

\textit{Notation}: We identify the set of $\R$-valued maps on $\N$ with
the topological vector space (hereafter TVS) $\R^\N$, with elements
of the form $x=(x^i)_{i\ge1}$. The set $\R^\N$ is endowed with its
canonical product topology, also called\vspace*{1pt} the topology of pointwise
convergence: $(x_n)_{n\ge1}$ in $\R^\N$ converges pointwise to $x\in
\R^\N$ if $x_n^i\to x^i$ for all $i\ge1$.
We set $\M=\R^{\N^2}$, whose elements are denoted by
$a=(a^{ij})_{i,j\ge1}$, define $\M_+$ as the subset of $\M$ composed
by elements with nonnegative components, and use the notation $\M
^1_+$ [resp., $\M_{f,+}$] to denote the set of elements $a$ in $\M_+$
such that $\sum_{i,j\ge1}a^{ij}<\infty$ [resp., only a finite number
of the $a^{ij}$'s are not equal to $0$].\vadjust{\goodbreak}

For\vspace*{1pt} $p \in[1,\infty)$ [resp., $p=\infty$], we denote by
$l^p$ [resp., $l^\infty$] the set of elements $x\in\R^\N$ such that
$|x|_{l^p}=( \sum_{i\ge1} |x^i|^{p})^{1/p}<\infty$ [resp.,
$|x|_{l^\infty}=\sup _{i\ge1} |x^i|<\infty$]. For the natural ordering,
$l_+^p$ is the closed cone of positive elements $x \in l^p$, that is,
$x^i \geq0$ for all $i$. Given\vspace*{1pt} $x,y\in\R^\N$, we write
$xy$ for $(x^1y^1,x^2y^2,\ldots)\in\R^\N$, $x/y$ for
$(x^1/y^1,x^2/y^2,\ldots)\in\R^\N$ and $x\cdot y$ for $\sum_{i\ge 1}
x^iy^i$ whenever it is well defined. To $j\in\N$, we associate the
element $e_j$ of $\R^\N$ satisfying $e_j^j=1$ and $e_j^i=0$ for $i \neq
j$. We shall also use the notation $\one=(1,1,\ldots)$.

We define $c_f$ as the space of finite real sequences, and $c_0$ as the
closed subspace of elements $x\in l^\infty$ such that $\lim_{i
\rightarrow\infty}x^i=0$. In the following, we\vspace*{1pt} shall use the notation
$\mu$ to denote an element of $(0,\infty)^\N$ such that $1/\mu\in
l^1$. To\vspace*{1pt} such a $\mu$, we associate the Banach space $l^1(\mu)$
[resp., the set $l^1_+(\mu)$] of elements $x\in\R^\N$ such that
$x\mu\in l^1$ [resp., $x\mu\in l^1_+$]. The Banach space $c_0(1/\mu)$
is defined accordingly. $x \in c_0(1/\mu)$ if and only if $x/\mu\in
c_0$. 
Recall that $l^1$ [resp., $l^1(\mu)$] is the topological dual of $c_0$
[resp., $c_0(1/\mu)$].

For a normed space $(E,\|\cdot\|_E)$, we define the natural distance
$d_{E}(x,y):=\|x-y\|_E$, denote by $d_{E}(x,A)$ [resp., $d_E(B,A)$] the
distance between $x$ [resp., the set $B\subset E$] and the set
$A\subset E$.

We shall work on a complete probability space $(\Omega,\Fc,\P)$
supporting a discrete-time filtration $\F=(\Fc_t)_{t\in\T}$. $\Fc
_0$ is the completion of the trivial $\sigma$-algebra and without loss
of generality, we assume that $\Fc_T=\Fc$.

Given a real locally convex TVS $E$, with topological dual $E'$, and a
$\sigma$-sub-algebra $\Gc\subset\Fc$, we denote by $E_w$ the linear
space $E$ endowed with the weak topology [i.e., the $\sigma(E,E')$
topology], $\Bc(E_w)$ stands for the corresponding Borel $\sigma
$-algebra, and we write $L^0(E,\Gc)$ to denote the collection of
weakly $\Gc$-measurable $E$-valued random variables. A subset $B$ of
$\Omega\times E$ is said to be weakly $\Gc$-measurable if $B \in\Gc
\otimes\Bc(E_w)$.
When \mbox{$(E,\|\cdot\|_E)$} is a separable Banach space, the elements of
$L^0(E,\Gc)$ are indeed strongly measurable; cf. Section V.4 of~\cite
{Yosida}. For $1\le p \le\infty$, we then use the standard notation
$L^p(E,\Gc)$ for the elements $X\in L^0(E,\Gc)$ such that $\mathbb
{E}[\|X\| _E^p]<\infty$ if $1\le p<\infty$, and $\|X\|_E$ is essentially
bounded if $p=\infty$. In the case of the nonseparable space
$l^\infty$, the elements $X \in L^0(l^\infty,\Gc)$ still have a $\Gc
$-measurable norm {$|X|_{l^\infty}$.} We therefore also use the
notation $L^p(l^\infty,\Gc)$ as defined above, although this space
does not have all the usual ``nice properties'' of $L^p$-spaces. We
omit $\Gc$ when $\Gc=\Fc$.

Any inequality between random variables or inclusion between random
sets has to be taken in the $\mbox{a.s.}$ sense.
%


%




\section{Model formulation} \label{S1}

\subsection{Financial strategies and no-arbitrage of second kind}

We consider a financial market in discrete time with proportional
transaction costs supporting a {countable infinite} number of tradable
assets. The evolution of the asset prices is described by a $(0,\infty
)^\N$-valued $\F$-adapted process $S=(S_t)_{t\in\T}$. Throughout
the paper, we shall impose the following technical condition:
%
%
\begin{equation}\label{hypborneprix}
S_t/S_s \in L^1(l^\infty) \qquad\mbox{for all } s, t\in\T.
\end{equation}
Similar conditions are satisfied in continuous time models without
transaction costs; cf. Theorem 2.2 of~\cite{ET}.
%
%
\begin{Remark}
Note that one could simply assume that
$S_t/S_s \in l^\infty$ for all $s, t\in\T$, which is a natural
condition, and replace the original measure $\P$ by $\tilde\P$
defined by
%
\[
d\tilde\P/d\P=\exp\biggl(-\sum_{s, t\in\T} |S_t/S_s|_{l^\infty
}\biggr)\Big/\mathbb{E}\biggl[\exp\biggl(-\sum_{s, t\in\T}
|S_t/S_s|_{l^\infty}\biggr)\biggr],
\]
which is equivalent and for which (\ref{hypborneprix}) holds.
\end{Remark}

The transaction costs are modeled as a $\M_+$-valued adapted process
$\lambda=(\lambda_t)_{t\le T}$. This means that buying one unit of
asset $j$ against units of asset $i$ at time $t$ costs $\pi
^{ij}_t:=(S^j_t/S^i_t)(1+\lambda^{ij}_t)$ units of asset $i$.

Throughout the paper, we shall assume that
%
%
\begin{equation} \label{eqinegatriangulairelambda}
\lambda^{ii}_t=0 \quad\mbox{and}\quad (1+\lambda^{ij}_t)(1+\lambda
^{jk}_t)\ge
(1+\lambda^{ik}_t) \qquad  \forall i,j,k\ge1 \mbox{ and }t
\in\T\hspace*{-28pt}
\end{equation}
and that
%
%
\begin{equation}\label{eqbornesuplambda}
{\sup_{t \in\T,i,j\ge1} }\|\lambda^{ij}_t \|_{L^\infty} < \infty.
\end{equation}
Note that these conditions have a natural economic interpretation. The
first is equivalent to $\pi^{ii}_t=1$ and $\pi^{ij}_t \pi^{jk}_t
\geq\pi^{ik}_t$ for all $i,j,k\ge1$ and $t\in\T$; compare with
\cite{schach}.

A portfolio strategy is described as a $\R^\N$-valued adapted process
$\xi=(\xi)_{t\in T}$ satisfying at any time $t \in\T$
\[
\xi^i_t\le\sum_{j\ge1}\bigl( a^{ji}-a^{ij}(1+\lambda
^{ij}_t)\bigr) \qquad \forall i\ge1\qquad \mbox{for some } a\in
L^0(\M_+,\Fc_t),
\]
whenever this makes sense, or equivalently,
%
%
\begin{equation} \label{eq1K}
-\xi_t\ge\sum_{i\neq j}a^{ij}\bigl((1+\lambda^{ij}_t)e_i
-e_j\bigr)\qquad \mbox{for some } a\in L^0(\M_+,\Fc_t).
\end{equation}

As explained in the \hyperref[secintroduction]{Introduction}, $\xi^i_t$
should be interpreted as
the additional net amount of money transferred at time $t$ to the
account invested in asset $i$ after making transactions on the
different assets. The quantity $a^{ji}$ should be interpreted as the
amount of money transferred to the account $i$ by selling
$a^{ji}(1+\lambda^{ji}_t)/S^j_t$ units of asset $j$. The above
inequality means that we allow the investor to throw away money from
the different accounts.\vadjust{\goodbreak}

In order to give a mathematical meaning to the above expressions, let
us define the random convex cones $\tilde K_t $ as the convex cones
generated by elements of finite length in $l_+^1$ and the set of
vectors on the right-hand side of (\ref{eq1K}) obtained by finite sums,
\begin{eqnarray*}
\tilde K_t(\omega)&=& \biggl\{x \in l^1\dvtx x = \sum_{i\neq j}a^{ij}\bigl(\bigl(1+
\lambda^{ij}_t(\omega)\bigr)e_i -e_j\bigr)+\sum_{i\ge1}b^ie_i \\
&&\hspace*{85pt}\mbox{
for some } a \in\M_{f,+}, b \in c_f \cap l_+^1 \biggr\}
\end{eqnarray*}
and define the set of admissible strategies as
\[
\Ac:=\{\xi=(\xi_t)_{t\in\T}\mbox{ $\F$-adapted}\dvtx\xi
_t\in-K_t \mbox{ for all } t \in\T\},
\]
where $K_t(\omega)$ denotes the $l^1$-closure of $\tilde K_t(\omega)$.
%
%
\begin{Remark}
Note that, by construction, $K_t(\omega)$ is a
closed convex cone in $l^1$ of vertex $0$ satisfying $l_+^1 \subset
K_t(\omega)$ and such that $K_t(\omega) \cap c_f$ is dense in
$K_t(\omega)$.
\end{Remark}

For ease of notation, we also define
\[
\Ac^T_{t}:=\{\xi\in\Ac\dvtx\xi_s=0 \mbox{ for } s<t\}.
\]

To an admissible strategy $\xi\in\Ac$, we associate the
corresponding portfolio process $V^\xi$ corresponding to a zero
initial endowment,
%
%
\begin{equation}\label{eqdefV}
V_t^{\xi}:= \sum_{s=0}^t \xi_sS_t/S_s.
\end{equation}
The $i$th component corresponds to the amount of money invested in the
$i$th asset at time $t$. Note that the additional amount of money $\xi
^i_s$ invested at time $s$ in the $i$th asset corresponds to $\xi
^i_s/S^i_s$ units of the $i$th asset, whose value at time $t$ is $(\xi
^i_s/S^i_s)S^i_t$.

We then define the corresponding sets of terminal portfolio values,
\[
\Xc^T_t:=\{V_T^\xi\dvtx\xi\in\Ac^T_t\}.
\]

We can now define our condition of no-arbitrage of the second kind,
which is similar to the one used in~\cite{DeKa10} and~\cite{ras09}
for finite-dimensional markets. It simply says that a trading strategy
cannot ensure that we end up with a solvable position at time $T$ if
the position was not already $\mbox{a.s.}$ solvent at previous times
$t\le T$.
%
%
\begin{Condition}[(NA2)] \label{CondNA2}
For all $t \in\T$, 
\[
\eta\in L^0(l^1,\Fc_t)\setminus L^0(K_t,\Fc_t) \quad\Rightarrow\quad
(\eta{S_T/S_t} +\Xc_t^T)\cap L^0(K_T) = \varnothing.
\]
\end{Condition}
%
%
\begin{Remark}\label{rempasKTsauf0}
For later use, note that it follows from \textbff{NA2} that $\Xc_0^T
\cap
L^0(K_T) =\{0\}$ whenever $K_t$ is $\mbox{a.s.}$ proper [i.e.,
$K_t\cap(-K_t)=\{0\}$]. Indeed, fix a nontrivial $\xi\in\Ac$ and
suppose that $V^\xi_T\in L^0(K_T)$. Since $\xi\neq0$, there is a
smallest $t^*$ such that $\xi_{t^*} \neq0$ (as a random variable). It
follows that $V^\xi_T=\xi_{t^*} {S_T/S_{t^*}} + g$ for some
$g\in\Xc_{t^*+1}^T$. The condition \textbff{NA2} then implies that
$\xi_{t^*} \in L^0(K_{t^*},\Fc_{t^*})$. However $\xi\in\Ac$, so
$\xi_{t^*}\in L^0(-K_{t^*},\Fc_{t^*})$. Since
$K_{t^*}\cap(-K_{t^*})=\{0\}$, this leads to a contradiction.
\end{Remark}
%
%
\begin{Remark}
Note that a simple condition implying \textbff{NA2} is: $\lambda$ is
constant (in time and $\omega$), and there exists a probability measure
$\Q\sim\P$ such that $S$ is a $\Q$-martingale. Indeed, under the above
assumption, $\eta{S_T/S_t}+ \sum_{s=t}^T \xi_sS_T/S_s \in K_{T}$
implies $\eta+\sum_{s=t}^T \E^{\Q}[\xi_s\mid\Fc_{t}] \in K_{T}$ by
convexity of $K_{T}$, for $\xi\in\Ac_{t}^{T}$. Since $\xi_{s}\in-K_{s}$
and the latter is constant and convex, we have
$\E^{\Q}[\xi_s\mid\Fc_{t}]\in-K_{s}=-K_{T}$. Hence, $\eta\in
K_{T}=K_{t}$.
\end{Remark}

\subsection{The efficient friction assumption}

In this paper, we shall assume that a version of the so-called
\textit{efficient friction} assumption holds. In finite-dimensional settings,
this means that $\lambda^{ij}_t+\lambda^{ji}_t>0$ for all $i\ne j$
and $t\in\T$, or equivalently that $K_t$ is a.s. proper [i.e.,
$K_t\cap(-K_t)=\{0\}$], or that the positive dual of each $K_t$ has
$\mbox{a.s.}$ nonempty interior, for all $t\in\T$; see~\cite{KSR01}.

In our infinite-dimensional setting, the positive dual cone of
$K_t(\omega)$ is defined as
\[
K^*_t(\omega) :=\{z\in l^\infty\dvtx z\cdot x \ge0 \mbox{ for
all }
x\in K_t(\omega)\}
\]
or, more explicitly,
%
%
\begin{equation} \label{K*}
K^*_t(\omega)=\bigl\{z\in l^\infty\dvtx0\le z^j\le z^i \bigl(1+\lambda
^{ij}_t(\omega)\bigr), i,j\ge1\bigr\},
\end{equation}
and the above mentioned condition could naively read
%
%
\begin{equation}\label{condefffrictionnaturelle}
\inf(\lambda^{ij}_t+\lambda^{ji}_t)>0,
\end{equation}
where the $\inf$ is taken over 
$t \in\T$ and $i \neq j$.
However, it is not sufficient in order to ensure that $K^*_t$ has
a.s. a nonempty interior, as shown in Remark~\ref{remlambdaij+ji} below.

We shall therefore appeal to a generalized version of the
\textit{Efficient Friction} (in short \textbff{EF}) assumption of
\cite{KSR01}
which is directly stated in terms of the random cones 
$K^*_t$ in $l^\infty$. Theorem~\ref{ThEF} below provides a natural
condition under which it is satisfied.
%
%
\begin{Assumption}[(EF)] \label{EffFric}
The $\M_+$-valued adapted process $\lambda$, satisfying (\ref
{eqinegatriangulairelambda}) and (\ref{eqbornesuplambda}), has the
property that for all $t\in\T$, and $\P$-a.e. $\omega$ the dual
cone $K^*_t(\omega)$ has an interior point $\theta_t(\omega)$ such
that $\theta_t \in L^0(l^\infty,\Fc_t)$.
\end{Assumption}

It is easy to find sufficient conditions on the transactions costs
$\lambda$ such that the Efficient Friction Assumption~\ref{EffFric}
is satisfied. The following result is a direct consequence of
Proposition~\ref{propEFNormCone} reported in Section~\ref{seclast} below.\vadjust{\goodbreak}

%
\begin{Theorem} \label{ThEF} Assume that
%
%
\begin{equation}\label{hypbornecout}
\inf\lambda^{ij}_t(\omega)>0\qquad \mbox{a.s.},
\end{equation}
where the $\inf$ is taken over 
$t \in\T$ and $i \neq j$.
Then the Efficient Friction Assumption~\ref{EffFric} is satisfied
with $\theta_t(\omega)=\one$.
\end{Theorem}
%
%
\begin{Remark} \label{remK*-linfty}
(1) If condition (\ref{hypbornecout}) is replaced by the weaker
one (\ref{condefffrictionnaturelle}) used in finite-dimensional
settings, then Theorem~\ref{ThEF} is no longer true. See Remark \ref
{remlambdaij+ji} for a counter example.\vspace*{-6pt}

\begin{longlist}[(2)]
\item[(2)] There are $\lambda$ giving rise to \textbff{EF} not
covered by Theorem~\ref{ThEF}. One such case is given by $\lambda$
defined by $\lambda^{ij}=1$ for all $i \neq j$ except $\lambda
^{12}=0$. In fact, for this case, Lemma~\ref{lmintK*} gives that
$(3/2,1,1,\ldots) \in\operatorname{int}(K^*_t)$.
\item[(3)] There are several possible generalizations of the concept
of \textit{Efficient Friction} to infinite-dimensional spaces. In
fact, in the finite-dimensional case a closed convex cone $C$ is proper
if and only if its dual cone
\[
C':=\{z\in E'\dvtx\sesq{z}{x}\ge0 \mbox{ for all } x \in C\}
\]
(we use here the notation $C'$ in place of $C^{*}$ since it is more
standard in the Banach space literature) has a nonempty interior,
while in the case of a Banach space we only have (see Section \ref
{seclast} for details)
\begin{eqnarray*}
&&\bigl(\operatorname{int}(C') \ne\varnothing\bigr)
\Rightarrow (C'
\mbox{ has the
generating property}) \\
&&\qquad\Leftrightarrow\quad
(C \mbox{ is normal}) \Rightarrow(C \mbox{ is proper}).
\end{eqnarray*}
So, in \textbff{EF} we have chosen the strongest of these conditions.
%
\item[(4)] Under \textbff{EF}, for all $\xi\in L^0(l^\infty,\Fc_t)$,
$d_{l^\infty}(\xi,\partial K^*_t)$ is a real $\Fc_t$-measurable
r.v., where $\partial K^*_t(\omega)$ is the border of $K^*_t(\omega
)$; see Section~\ref{seclast}. This is easy to prove, but nontrivial
since $l^\infty$ is not separable.
%
\item[(5)] The choice of the spaces has to be done with some care. For
instance, if the $\lambda^{ij}$'s are time independent and uniformly
bounded by some constant $c>0$, and if $\tilde K$ and $K$ are defined
in $l^p$ with $1< p < \infty$, instead of $l^1$, then $K^{*}=\{0\}$
and $K=l^p$. In fact, with $p^{-1}+ q^{-1}=1$, $y \in K^{*}$ if and
only if $y \in l^q$ and $0\le y^j\le y^i (1+\lambda^{ij})$ for all $i
\neq j\ge1$. In particular, $ \frac{y^j}{1+c} \le y^i$ for $i \neq
j\ge1$, so that $y \notin l^q$ whenever there exists $j\ge1$ such
that $y^j>0$. This shows that $K^{*}=\{0\}$, which then implies that $K=l^p$.
\end{longlist}
\end{Remark}

\section{Main results}\label{secmainresults}

In this section, we state our main results. The proofs are collected in
the subsequent sections.

From now on, we denote by $L_{t,b}^0$ the subset of random variables
$g\in L^0(l^1)$ bounded from below in the sense that
%
%
\begin{equation} \label{eqlbrv}
g+\eta S_T/S_t\in K_T \qquad\mbox{for some } \eta\in
L^0(l^1_+,\Fc_t).
\end{equation}

In the following, a subset $B \subset L_{t,b}^0$ is said to be
$t$-bounded from below if there exists $c \in L^0(\R_+,\Fc_t)$
(called a lower bound) such that any $g \in B$ satisfies (\ref{eqlbrv})
for some $\eta\in L^0(l^1_+,\Fc_t)$ such that $ |\eta
|_{l^1} \le c$.

Our first main result is a Fatou-type closure property for the sets
$\Xc_t^T$ in the following sense:
%
%
\begin{Definition}\label{deftFatouconv}
Let $(g_n)_{n\ge1}$ be a sequence in $L^0(l^1)$, which converges
$\mbox{a.s.}$ pointwise to some $g \in L^0(l^1)$ and fix $t\in\T$.

We say that $(g_n)_{n\ge1}$ is $t$-Fatou convergent with limit $g$ if
$\{g_n\dvtx n\ge1\}$ is a subset of $L_{t,b}^0$ which is $t$-bounded
from below.

We say that a subset $B$ of $L^0(l^1)$ is $t$-Fatou closed, if, for any
sequence $(g_n)_{n\ge1}$ in $B$, which $t$-Fatou converges to some
$g\in L^0(l^1)$, we have $g\in B$.
\end{Definition}
%
%
\begin{Theorem}\label{thmSTFatouclosure} Assume that \textup{\textbff{NA2}} and
\textup{\textbff{EF}} hold. Then $\Xc_t^T$ is $t$-Fatou closed, for all $t\in\T$.
\end{Theorem}
%
%
\begin{Remark}\label{remcounterex}
We shall provide in Section~\ref{subcontexemplefermeture} a counter
example showing that the above closure property does not hold in
general if we replace Assumption~\ref{EffFric} by the weaker one,
$\lambda^{ij}_t + \lambda^{ji}_t >0$ for all $t\in\T$ and $i\neq j$.
The question whether it holds under (\ref{condefffrictionnaturelle})
above is left open.
\end{Remark}

The above Fatou closure property can then be translated in a $*$-weak
closure property of the set of terminal portfolio holding labeled in
time-$t$ values of the assets, that is, $S_t \Xc_t^T/S_T=\{S_tV_T/S_T,
V_T\in\Xc_t^T\}$. Recall that $\mu$ denotes any element of $\R
^\N$ such that $1/\mu\in l^1_+$.
%
%
\begin{Theorem}\label{thmfermeturefaible} Assume that \textup{\textbff{NA2}} and
\textup{\textbff{EF}} hold. Then, the set $(S_t \Xc_t^T/\break S_T) \cap L^\infty
(l^1(\mu))$ is $\sigma(L^\infty(l^1(\mu))$, $L^1(c_{0}(1/\mu
)))$-closed for all $t \in\T$.
\end{Theorem}
%
%
\begin{Remark}\label{remmuarbitraire}
Note that we use the spaces $l^1(\mu)$ and $c_{0}(1/\mu)$, with
$\mu\in(0,\infty)^\N$ such that $1/\mu\in l^1$, in the above
formulation instead of the more natural ones $l^1$ and $c_0$. The
reason is that bounded sequences $(x_n)_{n\ge1}$ in $l^1(\mu)$ have
components satisfying $|x^i_n| \le c 1/\mu^i$ for some $c>0$
independent of $i$ and $n$ and where $1/\mu\in l^1_+$. In particular,
$x+c/\mu\in l^1_+$. This allows us to appeal to the Fatou closure
property of Theorem~\ref{thmSTFatouclosure}; see the proof of Theorem
\ref{thmfermeturefaible} in Section~\ref{secclosure}. We shall actually
see in Remark~\ref{remfermpaspossibledansl1} below that the above
closure property cannot be true in general if we consider the (more
natural) $\sigma(L^\infty(l^1)$, $L^1(c_{0}))$-topology.
\end{Remark}

By using standard separation arguments, Theorem \ref
{thmfermeturefaible} allows us, as usual, to characterize the set of attainable
claims in terms of
natural dual processes.

In models with proportional transaction costs, they consist of elements
of the sets $\Mc_{{t}}^T(K^*\setminus\{0\})$ of $\R^{\N}$-valued
$\F$-adapted processes $Z$ on $\T_t:=\{t,t+1, \ldots,T\}$ such that
$Z_s\in K^*_s\setminus\{0\}$, for all $s \in\T_t$, and $ZS$ is a $\R
^{\N}$-valued martingale on~$\T_t$, \mbox{$t\in\T$}. Following the
terminology of~\cite{schach}, elements of the form $ZS$ with $Z\in\Mc
_{{t}}^T(K^*\setminus\{0\})$ are called consistent price system (on
$\T_t$).
%
%
\begin{Theorem}\label{thmsuperhedging} Assume that \textup{\textbff{NA2}} and
\textup{\textbff{EF}} hold. Fix $t\in\T$. Then, $\Mc_t^T(K^*\setminus\{0\})\ne
\varnothing$. Moreover, for any $g\in L^0(l^1)$ such that $g+\eta
S_T/S_t \in L^0(l^1_+)$ for some $\eta\in L^0(l^1_+, \Fc_t)$, we have
\[
g \in\Xc_{t}^T \quad\Leftrightarrow\quad \mathbb{E}[Z_T\cdot g\mid\Fc_t]\le0
\qquad\mbox{for all } Z\in\Mc_t^T(K^*\setminus\{0\}).
\]
\end{Theorem}

We note that the above conditional expectation $\mathbb{E}[Z_T\cdot
g\mid\Fc_t]$ is well defined as a $\R\cup\{\infty\}{}$-valued $\Fc
_t$-measurable r.v. In fact $g+\eta S_T/S_t \in L^0(l^1_+)$ implies
that $Z_T \cdot g \geq- Z_T \cdot(\eta S_T/S_t)$ where $\eta/S_t\in
L^0(l^1,\Fc_t)$ and $Z_T S_T\in L^1(l^\infty)$ by definition.

Following arguments used in~\cite{ras09} and~\cite{DeKa10}, one can
also prove that the so-called \textbff{B} condition holds under
\textbff{NA2}.
%
%
\begin{Condition}[(B)] \label{assB} The following holds for all $t\in
\T$ and $\xi\in L^0(l^1,\Fc_t)$:
\[
Z_t\cdot\xi\ge0\qquad \forall Z\in\Mc_t^T(K^*\setminus\{0\})
\Rightarrow\xi\in K_t.
\]
\end{Condition}
%
%
\begin{Theorem} \label{thmNA2equivaB}
\textup{\textbff{NA2}}${}\Leftrightarrow{}$(\textup{\textbff{B}} and ${\Mc
}_0^T(K^*\setminus\{0\})\ne\varnothing$).
\end{Theorem}

It finally\vspace*{1pt} implies the existence of Strictly Consistent Price Systems,
that is, elements of the sets $\Mc_t^T(\operatorname{int}{K^*})$ of processes
$Z\in\Mc_t^T(K^*\setminus\{0\})$ such that $Z_s\in\operatorname{int}{K^*_s}
$, for all $s \in\T_t$. The \textbff{NA2} condition actually turns out to
be equivalent to the existence of a sufficiently big sets of consistent
price systems, which is referred to as the Many Consistent Price
Systems (\textbff{MCPS}) and Many Strictly Consistent Price Systems
(\textbff{MSCPS}) properties.
%
%
\begin{Condition}\label{condMCPS}
We say that the condition \textbff{MCPS} [resp., \textbff{MSCPS}]
holds if
for all $t \in\T$ and $\eta\in L^0(\operatorname{int}{K^*_t},\Fc_{t})$ such
that $\eta S_t\in L^1(l^\infty,\Fc_t)$, there exists $Z\in\Mc
_t^T(K^*\setminus\{0\})$ [resp., $Z\in\Mc_t^T(\operatorname{int}K^*)$] such
that $Z_t=\eta$.
\end{Condition}
%
%
\begin{Theorem} \label{thNA2equivMCPSeuivMSCPS} Assume that
\textup{\textbff{EF}}
holds. Then,
the three conditions \textup{\textbff{NA2}}, \textup{\textbff{MCPS}} and
\textup{\textbff{MSCPS}} are
equivalent.
\end{Theorem}

\section{Closure properties and duality}\label{secclosure}

We start with the proof of our closure properties which are the main
results of this paper.

\subsection{Efficient frictions and Fatou closure property}

The key idea for proving the closure property of Theorem \ref
{thmSTFatouclosure} is the following direct consequence of the \textbff{EF}
Assumption~\ref{EffFric}.\vadjust{\goodbreak}
%
%
\begin{Corollary}\label{corbornel1siborneinfell}
Suppose that \textup{\textbff{EF}} holds. Then, for all $t\in\T$, there exists
$\alpha\in L^0(\R_+,\Fc_t)$ such that
\[
|\xi|_{l^1}\le\alpha|\eta|_{l^1}\qquad \forall(\xi,\eta) \in
L^0(-K_t,\Fc_t)\times L^0(K_t,\Fc_t) \qquad\mbox{such that }
\xi
+\eta\in K_t.
\]
\end{Corollary}
\begin{pf}
According to the \textbff{EF} Assumption~\ref{EffFric} there exists
$\theta_t \in L^0(l^\infty,\Fc_t)$ such that $\theta_t(\omega)$ is
an interior point of $K^*_t(\omega)$ for $\P$-a.e. $\omega\in\Omega
$. Define
\[
\alpha(\omega):=8 |\theta_t(\omega)|_{{l^\infty}} \biggl(\frac
{1}{d_{l^\infty}(\theta_t(\omega),\partial K_t'(\omega))}\biggr)^2.
\]
Then $\alpha\in L^0(\R_+,\Fc_t)$ by (4) of Remark~\ref{remK*-linfty}.
We observe that $\xi_t(\omega) \in(K_t(\omega)-\eta
_t(\omega)) \cap(\eta_t(\omega)-K_t(\omega))$, according to the
hypotheses and the fact that $K_t+K_t=K_t$. Lemma~\ref{lemknormcone}
and Lemma~\ref{lembornel1siborneinfell2}, with
$C=K_t(\omega)$, $f_0=\theta_t(\omega)$, $x=\xi_t(\omega)$,
$y=\eta_t(\omega)$ and $b=1/2$, then apply, which proves the
corollary with the above defined $\alpha$.
\end{pf}

As an almost immediate {consequence} of the above corollary, we can now
obtain under \textbff{NA2} the following important property of sequential
relative compactness of lower bounded subsets {[see (\ref{eqlbrv})]} of
\[
\Xc_{t,b}^T:=\Xc_t^T\cap L_{t,b}^0.
\]

%
\begin{Corollary}\label{coroxinbornedansl1} Assume that
\textup{\textbff{EF}} and \textup{\textbff{NA2}} hold. Fix $t\in\T$ and
let $(\xi^n)_{n\ge1}$ be a sequence in $\Ac_t^T$ such that
$(V^{\xi^n}_T)_{n\ge1}$ is a sequence in $\Xc_{t,b}^T$ which is
t-bounded from below. Then:

\begin{longlist}
\item
$(\xi^n_t)_{n\ge1}$ is $\mbox{a.s.}$ bounded in $l^1$.

\item There is a sequence $(n_k)_{k\ge1}$ in $L^0(\N,\Fc_t)$
such that $(\xi^{n_k}_t)_{k\ge1}$ converges pointwise $\mbox{a.s.}$
to some
$\xi_t\in L^0(-K_t,\Fc_t)$.
\end{longlist}
\end{Corollary}
\begin{pf}
Let $c \in L^0(\R_+,\Fc_t)$ be a lower bound for $(V^{\xi
^n}_T)_{n\ge1}$ so that $(V^{\xi^n}_T,\eta_n)$ satisfy (\ref{eqlbrv})
in place of $(g,\eta)$, for all $n\ge1$, where the sequence
$(\eta_n)_{n\ge1}$ in $L^0(l^1_+,\Fc_t)$ satisfies $\sup_{n \ge1}
|\eta_n|_{l^1} \le c$.

\begin{longlist}
\item
We then have $V^{\xi^n}_T+\eta_n S_T/S_t = (\eta_n +\xi
^n_t)S_T/S_t +(V^{\xi^n}_T-\xi^n_tS_T/S_t)\in K_T$ where $V^{\xi
^n}_T-\xi^n_tS_T/S_t \in\Xc_{t+1}^T$, recall (\ref{eqdefV}).
Hence, \textbff{NA2} implies that $ \eta_n +\xi^n_t\in K_t$. The claim
then follows from Corollary~\ref{corbornel1siborneinfell},
$l^1_+\subset K_t$ and the fact that $\sup_{n \ge1} |\eta_n|_{l^1}
\le c$, which imply $\sup_{n \ge1} |\xi^n_t|_{l^1} \le\alpha c$ for
some $\alpha\in L^0(\R_+,\Fc_t)$.

\item It follows, in particular from the above claim, that $|(\xi
^{n}_t)^i| \le\alpha c$ for all $n,i\ge1$. For $i=1$, we can then
construct a $\Fc_t$-measurable sequence $(n^1_k)_{k\ge1}\in L^0(\N
,\Fc_t)$ such that
$((\xi^{n^1_k}_t)^1)_{k\ge1}$ converges $\mbox{a.s.}$ and is also
$\mbox{a.s.}$
uniformly bounded in $l^1$; see, for example,~\cite{teachersnote}.
Iterating this procedure on the different components, we obtain after
$\kappa$ steps a sequence $(n^\kappa_k)_{k\ge1} \in L^0(\N,\Fc_t)$
such that
$((\xi^{n^\kappa_k}_t)^i)_{k\ge1}$ converges $\mbox{a.s.}$ for all
$i\le
\kappa$. It follows that the sequence $(\xi^{n^k_k}_t)_{k\ge1}$
converges $\mbox{a.s.}$ pointwise to some $\Fc_t$-measurable random variable
$\xi_t$ with values in $\R^\N$. Since $|\xi^n_t|_{l^1}$ is $\mbox{a.s.}$
uniformly bounded, $\xi_t \in l^1$ a.s.\qed
\end{longlist}
\noqed\end{pf}

We can now conclude the proof of Theorem~\ref{thmSTFatouclosure} by
appealing to an inductive argument.
\begin{pf*}{Proof of Theorem~\ref{thmSTFatouclosure}}
If $t=T$, the result is an immediate consequence of Corollary \ref
{coroxinbornedansl1}. We now assume that it holds for some $0<t+1\le T$
and show that this implies that it holds for $t$ as well. Let
$(g_n)_{n\ge1}$ be a sequence in $\Xc_t^T$ which is $t$-Fatou
convergent with limit $g \in L^0(l^1)$. Then, by definition, there
exist $c \in L^0(\mathbb{R},\Fc_t)$ and $\eta_n \in L^0(l^1_{+},\Fc_t)$
such that $|\eta_n|_{l^1} \le c$ and $g_n+\eta_n S_T/S_t \in K_T$ for
all $n\ge1$. Let the sequence $(\xi^n)_{n\ge1}$ in $\Ac_t^T$ be such
that $V^{n}_T=g_n$ for all $n\ge1$, where $V^n=V^{\xi^n}$. It then
follows from Corollary~\ref{coroxinbornedansl1} that we can find a
sequence $(n_k)_{k\ge1}$ in $L^0(\N,\Fc_t)$ such that
$(\xi^{n_k}_t)_{k\ge1}$ is $\mbox{a.s.}$ bounded in $l^1$ and converges
pointwise $\mbox{a.s.}$ to some $\xi_t\in L^0(-K_t,\Fc_t)$. Clearly,
$(\xi^{n_k})_{k\ge1}$ is a sequence in $\Ac_t^T$ since $(n_k)_{k\ge
1}$ is $\Fc_t$-measurable, and $V^{n_k}_T=g_{n_k}$ where the later
converges $\mbox{a.s.}$ pointwise to $g$ as \mbox{$k\to\infty$}. Moreover, $
g_{n_k}- \xi^{n_k}_t S_T/S_t= V^{n_k}_T -\xi^{n_k}_t S_T/S_t\in
\Xc_{t+1}^T$ and $ (g_{n_k}- \xi^{n_k}_tS_T/S_t) +(\eta_{n_k}+
\xi^{n_k}_t) S_T/S_t \in L^0(K_T)$. Since $(\eta_{n_k}+
\xi^{n_k}_t)_{k\ge1}$ is $\mbox{a.s.}$ bounded in $l^1$ and $ (g_{n_k}-
\xi^{n_k}_tS_T/S_t)_{k\ge1}$ converges $\mbox{a.s.}$ pointwise to
$g- \xi_t
{S_T/S_t}\in\Xc_{t+1}^T$, the fact that $\Xc_{t+1}^T$ is $(t+
1)$-Fatou closed, this implies that $g- \xi_t {S_T/S_t}\in\Xc_{t+
1}^T$ and therefore that $g\in\Xc_{t}^T$.
\end{pf*}

\subsection{Weak closure property and the dual representation of
attainable claims}

We now turn to the proof of Theorem~\ref{thmfermeturefaible} which
will allow us to deduce the dual representation of Theorem \ref
{thmsuperhedging} by standard separation arguments. It is an easy
consequence of Theorem~\ref{thmSTFatouclosure} once the suitable
spaces have been chosen.
\begin{pf*}{Proof of Theorem~\ref{thmfermeturefaible}}
Fix $t \in\T$ and set $F=L^1(c_{0}(1/\mu))$, so that $F'=L^\infty
(l^1(\mu))$, where we recall that 
$1/\mu\in l^1_{+}$.
Let $B_1$ denote the unit ball in $F'$, and define the set $\Theta:=
(S_t \Xc_t^T/S_T) \cap B_1$.

By the Krein--\v{S}mulian theorem (cf. corollary, Chapter IV, Section
$6.4$ of~\cite{Schaefer}), it suffices to show that $\Theta$ is
$\sigma(F',F)$-closed. To see this, let $(h_\alpha)_{\alpha\in\Ic
}$ be a net in $\Theta$ which converges $\sigma(F',F)$ to some $h\in
B_1$. After possibly passing to convex combinations, we can then
construct a sequence $(f_n)_{n\ge1}$ in $\Theta$ which convergences
$\mbox{a.s.}$ pointwise to $h$. In fact, this follows from Lemma~\ref
{convexclosure} below with $E=(L^1(\R))^\N$.
This implies that the sequence $(f_nS_T/S_t)_{n\ge1}$ in $\Xc_t^T$
converges to $h S_T/S_t$ a.s. pointwise. Since $f_n \in B_1$, we
have $f_n+ {1/\mu}\in l^1_+$, and therefore $f_n S_T/S_t+ {(1/\mu)}
S_T/S_t \in K_T$. This shows that the sequence $(f_nS_T/S_t)_{n\ge1}$
is $t$-Fatou convergent with limit $hS_T/S_t\in L^0(l^1)$. It thus
follows from Theorem~\ref{thmSTFatouclosure} that $hS_T/S_t\in\Xc
_t^T$ and therefore that $h\in\Theta$.\vadjust{\goodbreak}
\end{pf*}

To complete the proof of Theorem~\ref{thmfermeturefaible}, we now
state the following technical lemma which was used in the above arguments.
%
%
\begin{Lemma} \label{convexclosure}
Let $E$ and $F$ be locally convex TVS, 
with topological duals $E'$ and $F'$ and let $\mathfrak{T}(E)$ be the
topology of $E$. Suppose $F' \subset E$, $E' \subset F$ and that $E$ is
metrizable.
If $(x_\alpha)_{\alpha\in\Ic}$ is a net in $F'$, with convex hull
$J$ and converging in the $\sigma(F',F)$ topology to $x$, then there
exists a sequence $(y_n)_{n\ge1}$ in $J$, which is $\mathfrak{T}(E)$
convergent to $x$.
\end{Lemma}
\begin{pf}
Since $F' \subset E$ and $E' \subset F$, the topology on $F'$ induced
by $\sigma(E,E')$ is weaker than the $\sigma(F',F)$ topology. The net
$(x_\alpha)_{\alpha\in\Ic}$ then also converges in the $\sigma(E,E')$
topology, so $x \in\bar J$ the $\sigma(E,E')$-closure of $J$. Since
$\bar J$ is also $\mathfrak{T}(E)$-closed (cf. Corollary 2, Chapter II,
Section 9.2 of~\cite{Schaefer}) and $(E, \mathfrak{T}(E))$ is
metrizable, it now follows that there exists a sequence in $J$ which is
$\mathfrak{T}(E)$-convergent to~$x$.
\end{pf}

From now on, we follow the usual ideas based on the Hahn--Banach
separation theorem. For ease of notation, we set $\tilde\Xc_0^T =
(S_0 \Xc_0^T/S_T) \cap L^\infty(l^1(\mu))$, and let $\tilde\Xc
_{s,0}^T$ denote the set of elements of the form $-\alpha e_i
S^i_0/S^i_t\chi_{\{S^i_t\ge\eps\}} $ or $\alpha(e_j-(1+\lambda
^{ij}_t) e_i) S_0/S_t\chi_{\{S^j_t\wedge S^i_t\ge\eps\}}$ for some
$t\in\T$, $i,j\ge1$, $\eps>0$ and $\alpha\in L^\infty(\R_+,\Fc
_t)$. Note that
%
%
\begin{equation}\label{eqinclusiontildeXc}
\tilde\Xc_{s,0}^T\subset\tilde\Xc_{0}^T.
\end{equation}

%
%
\begin{Proposition}\label{propsepa1}
\textup{(1)}
Suppose that \textup{\textbff{EF}} and \textup{\textbff{NA2}} hold. Then, for all $\eta
\in L^\infty(l^1(\mu)) \setminus\tilde\Xc_0^T$, there exists $Y
\in L^1(c_{0}(1/\mu))$ such that
\[
\mathbb{E}[Y\cdot X]\le0 <\mathbb{E}[Y\cdot\eta] \qquad\mbox{for all
} X\in\tilde\Xc
_0^T.
\]

\textup{(2)} Suppose that $0 \neq Y \in L^1(c_{0}(1/\mu))$ and that for
all $X\in\tilde\Xc_{s,0}^T$
\[
\mathbb{E}[Y\cdot X]\le0.
\]
Then $ Z_t:=\mathbb{E}[Y\mid\Fc_t]S_0/S_t$ satisfies $Z_tS_t=\mathbb
{E}[S_T Z_T \mid\Fc_t]$ and $Z_t \in L^0(K^*_t,\Fc_t)\setminus\{0\}$
for all $t
\in\T$.
%
\end{Proposition}
\begin{pf}
In this proof, we use the notation $F:=L^1(c_{0}(1/\mu))$ and
$F':=L^\infty(l^1(\mu))$.\vspace*{8pt}

(1) The set $\tilde\Xc_0^T$ being convex and $\sigma(F',F)$-closed,
by Theorem~\ref{thmfermeturefaible}, it follows from the
Hahn--Banach separation theorem that we can find $Y \in F$ such that
\[
\sup_{X\in\tilde\Xc_0^T}\mathbb{E}[Y \cdot X] < \mathbb{E}[Y
\cdot\eta].
\]
Since $\tilde\Xc_0^T$ is a cone that contains $0$, we clearly have
%
%
\begin{equation}\label{eqinegaseparation}
\sup_{X\in\tilde\Xc_0^T}\mathbb{E}[Y \cdot X] =0< \mathbb{E}[Y
\cdot\eta].\vadjust{\goodbreak}
\end{equation}

(2) First note that $\mathbb{E}[Y \mid\Fc_t] \in F$, so that $Z$ is
well defined as a $\R^\N$-valued process, and that (\ref
{eqinegaseparation}) implies $Z_T \neq0$ as a random variable.
Moreover, the fact that the left-hand side inequality of the
proposition holds for simple strategies of the form $-\alpha e_i
S^i_0/S^i_t\chi_{\{S^i_t\ge\eps\}} $ and $\alpha(e_j-(1+\lambda
^{ij}_t) e_i) S_0/S_t\chi_{\{S^j_t\wedge S^i_t\ge\eps\}}$, for all
$t\in\T$, $i,j\ge1$, $\eps>0$ and $\alpha\in L^\infty(\R_+,\Fc
_t)$, implies that $Z_t:=\mathbb{E}[Y\mid\Fc_t]S_0/S_t= \mathbb
{E}[S_T Z_T \mid\Fc_t]/S_t $ satisfies $0\le Z_t^j \le Z_t^i
(1+\lambda^{ij}_t)$, $i,j\ge
1$, for all $t\in\T$. Hence, $Z_t \in K^*_t$ by (\ref{K*}). Finally,
$\mathbb{P}[{Z=Z_T \neq0}]>0$ implies that $\mathbb{P}[{Z_t \neq
0}]>0$ for $t< T$.
\end{pf}
%
%
\begin{Remark}\label{remfermpaspossibledansl1}
Note that the statement of Theorem~\ref{thmfermeturefaible} cannot be
true in general if we consider the weak topology $\sigma(L^\infty
(l^1),L^1(c_0))$ on the space $(S_t \Xc_t^T/S_T) \cap L^\infty(l^1)$
instead of $\sigma(L^\infty(l^1(\mu))$, $L^1(c_0(1/\mu)))$ on $(S_t
\Xc_t^T/S_T) \cap L^\infty(l^1(\mu))$. Indeed, if $S_0\Xc_t^T/S_T\cap
L^\infty(l^1) $ was closed in the topology
$\sigma(L^\infty(l^1),\break L^1(c_0))$, then the same arguments as in the
proof of Proposition~\ref{propsepa1} above would imply the existence of
a random variable $Z_T$ such that $Z_T\in K^*_T\setminus\{0\}$ and
$Z_TS_T/S_0 \in c_0$. Recalling (\ref{K*}), this would imply that $0\le
Z^j_T\le(1+\lambda^{ij}_T)Z^i_T$ for all $i,j\ge1$ and
$Z^i_TS^i_T/S^i_0\to0$ a.s. as $i\to\infty$. Since $Z^1_T$ is not
identically equal to $0$, this cannot hold, except if $S^i_T/S^i_0\to
0$ as $i\to\infty$ on a set of nonzero measure, which is in
contradiction with (\ref{hypborneprix}). The closure property stated in
terms of $\sigma(L^\infty(l^1(\mu)),L^1(c_0(1/\mu)))$ does obviously
not lead to such a contradiction since (\ref{eqbornesuplambda}) and
(\ref{hypborneprix}) imply that $Z_T S_T/S_0\in l^\infty$ so that
$(Z^i_TS^i_T/S^i_0)/\mu^i \to0$ a.s. as $i\to\infty$, whenever
$1/\mu\in l^1$.
\end{Remark}
%
%
\begin{Corollary}\label{coroMtTnonvide} Suppose that \textup{\textbff{EF}}
and \textup{\textbff{NA2}} hold. Then, $\Mc_t^T(K^*\setminus\{0\})\ne
\varnothing$ for all $t\in\T$.
\end{Corollary}
\begin{pf}
It follows from \textbff{NA2} that $e_1\in L^\infty(l^1(\mu))\setminus
\tilde\Xc_0^T$. Using Proposition~\ref{propsepa1} and (\ref
{eqinclusiontildeXc}) then implies that there exists $Y \in
L^1(c_{0}(1/\mu))$ such that
%
%
\begin{equation}\label{eqinequaYc}
\mathbb{E}[Y\cdot X]\le0 <\mathbb{E}[Y\cdot e_1] \qquad\mbox{for all }
X\in\tilde\Xc
_{s,0}^T.
\end{equation}
Let $\Yc$ denote the set of random variables $Y\in L^1(c_{0}(1/\mu))$
satisfying the left-hand side of (\ref{eqinequaYc}) for all $ X\in
\tilde\Xc_{s,0}^T$. We claim that there exists $\tilde Y\in\Yc$
such that $a:=\sup_{Y\in\Yc}\mathbb{P}[{Y^1>0}]=\mathbb{P}[{\tilde
Y^1>0}]$. To
see this, let $(Y_n)_{n\ge1}$ be a maximizing sequence. It follows
from Proposition~\ref{propsepa1} that $\mathbb{E}[Y_n]\in K^*_0$ and
$Y^i_n\ge0$ for all $i\ge1$. Moreover, we can assume that $\mathbb
{P}[{Y^1_n>0}]>0$. We can then choose $(Y_n)_{n\ge1}$ such that
$\mathbb{E}[Y^1_n]=1$.
Recalling (\ref{eqbornesuplambda})--(\ref{K*}), this implies that
there exists $c>0$ such that $0\le\mathbb{E}[Y^i_n]\le(1+c) \mathbb
{E}[Y^1_n]=(1+c)$ for all $i\ge1$. Using Komlos lemma, a diagonalization
argument and Fatou's lemma, we can then assume, after possibly passing
to convex combinations, that $(Y_n)_{n\ge1}$ converges $\mbox{a.s.}$
pointwise to some $Y\in L^1(\R_+)^\N$. Set $\tilde Y:=\sum_{n\ge1}
2^{-n} Y_n$. It follows from the monotone convergence theorem that it
satisfies the left-hand side of (\ref{eqinequaYc}) for all $X\in
\tilde\Xc_{s,0}^T$. Moreover, $\mathbb{P}[{\tilde Y^1>0}]\ge\mathbb
{P}[{Y^1_n>0}]\to a$ so that $\mathbb{P}[{\tilde Y^1>0}]=a$. We now
show that $\mathbb{P}[{\tilde Y^1>0}]=1$. If not, there exists $A\in
\Fc$ with $\mathbb{P}[{A}]>0$
such that $\tilde Y^1=0$ on $A$. Since $e_1\chi_A \in L^\infty
(l^1(\mu))\setminus\tilde\Xc_0^T$, by \textbff{NA2}, it follows from
Proposition~\ref{propsepa1} that we can find $Y \in L^1(c_{0}(1/\mu
))$ such that
such that
\[
\mathbb{E}[Y\cdot X]\le0 <\mathbb{E}[Y\cdot e_1 \chi_A]
\qquad\mbox{for all } X\in\tilde
\Xc_{0}^T.
\]
By (\ref{eqinclusiontildeXc}), $Y+\tilde Y \in\Yc$ and $\mathbb
{P}[{Y^1 +\tilde Y^1>0}]>\mathbb{P}[{\tilde Y^1>0}]$ since $\mathbb
{E}[Y\cdot e_1 \chi_A]>0$
implies that $\mathbb{P}[{\{Y^1 >0\}\cap A}]>0$, a contradiction. To conclude
the proof it suffices to observe that $Z$ defined by $ \tilde
Z_t:=\mathbb{E}[\tilde Y\mid\Fc_t]S_0/S_t$ satisfies $\tilde
Z_tS_t=\mathbb{E}[S_T \tilde Z_T \mid\Fc_t]$ and $\tilde Z_t \in
L^0(K^*_t,\Fc_t)\setminus\{0\}$
for all $t \in\T$, by Proposition~\ref{propsepa1} again. Moreover,
(\ref{K*}) and $\mathbb{P}[{\tilde Y^1>0}]=1$ implies that $\mathbb
{P}[{\tilde Y^i>0}]=1$ for all $i\ge1$. This shows that $\tilde Z_t
\in
L^0(K^*_t\setminus\{0\},\Fc_t)$ for all $t \in\T$.
\end{pf}

The statement of Theorem~\ref{thmsuperhedging} is then deduced from
Proposition~\ref{propsepa1} and the following standard result.
%
%
\begin{Lemma}\label{rmsupmartproperty} Fix $\xi\in\Ac_t^T$ and
$Z \in\Mc_t^T(K^*\setminus\{0\})$, for some $t\in\T$. If $V^\xi
_T+\eta S_T/S_t \in K_T$ for some $\eta\in L^0(l^1,\Fc_t)$, then
\[
Z_s\cdot V^\xi_{s-1}S_{s}/S_{s-1}\ge Z_s\cdot V^\xi_s\ge\mathbb
{E}\bigl[Z_{(s+1)\wedge T} \cdot V^\xi_{(s+1)\wedge T}\mid\Fc_s\bigr]\ge-Z_s
\cdot
\eta S_s/S_t
\]
for all $t\le s \le T$,
with the convention $V^\xi_{-1}/S_{-1}=0$.
\end{Lemma}
\begin{pf}
Note that the left-hand side inequality just\vspace*{1pt} follows from
the fact that $\xi_s\in-K_s$ while $Z_s \in K^*_s$, and the definition
of $V^\xi$ in (\ref{eqdefV}). We now prove the two other inequalities.
For $s=T$, it follows from the fact that $Z_T\in K^*_T$ and $V^\xi
_T+\eta S_T/S_t \in K_T$. Assuming that it holds for $t<s+1\le T$, we
have $Z_{s+1}\cdot V^\xi_{s+1}\ge- Z_{s+1} \cdot\eta S_{s+1}/S_t$. On
the other hand, the already proved, left-hand side inequality above
implies $Z_{s+1} \cdot V^\xi_{s+1} \le Z_{s+1} \cdot V^\xi
_{s}S_{s+1}/S_s$. Since $\mathbb{E}[Z_{s+1}S_{s+1} \mid\Fc_s]=Z_{s
}S_{s } $ by definition of $\Mc_t^T(K^*\setminus\{0\})$, this shows
that the above property holds for $s$ as well.
\end{pf}

We now turn to the proof of Theorem~\ref{thmsuperhedging}. The basic
argument is standard, up to additional technical difficulties related
to our infinite-dimensional setting.
\begin{pf*}{Proof of Theorem~\ref{thmsuperhedging}}
The fact that $\Mc_t^T(K^*\setminus\{0\})\ne\varnothing$ for all
$t\in
\T$ follows from Corollary~\ref{coroMtTnonvide}. We now fix $g \in
L_{t,b}^0$. In view of Lemma~\ref{rmsupmartproperty}, it is clear
that
\[
g \in\Xc_{t}^T \Rightarrow\mathbb{E}[Z_T\cdot g\mid\Fc_t]\le0
\qquad\mbox{for all } Z\in\Mc_t^T(K^*\setminus\{0\}).
\]
It remains to prove the converse implication. We therefore assume that
%
%
\begin{equation}\label{eqpreuvecompliquee}
\mathbb{E}[Z_T\cdot g\mid\Fc_t]\le0 \qquad\mbox{for all } Z\in\Mc
_t^T(K^*\setminus\{0\})
\end{equation}
and show that $g\in\Xc_t^T$.

\begin{longlist}
\item
The case where $S_0g/S_T \in L^\infty(l^1(\mu))$ is handled
by very standard arguments based on Proposition~\ref{propsepa1} and
Corollary~\ref{coroMtTnonvide}. We omit the proof.\vadjust{\goodbreak}

\item We now turn to the case where $g \in L^0(l^1(\mu))$ is such
that $g+\eta S_T/S_t \in K_T$ for some $\eta\in L^0(l^1_+(\mu),\Fc_t)$.
We first construct a sequence $(g_n)_{n\ge1}$ defined as
$g_n:=(g\one_{\{|S_0g/S_T|_{l^1(\mu)}\le n\}}- \eta(S_T/S_t)\one_{\{
|S_0g/S_T|_{l^1(\mu)}>n\}})\one_{\{|S_0\eta/S_t|_{l^1(\mu)}\le n\}
}$. Since (\ref{eqpreuvecompliquee}) holds, $g-g_n\in K_T$ on $\{
|S_0\eta/S_t|_{l^1(\mu)}\le n\}\in\Fc_t$ and $Z_T\in K^*_T$ for
$Z\in\Mc_t^T(K^*\setminus\{0\})$, we have $ \mathbb{E}[Z_T\cdot
g_n\mid\Fc_t]\one_{\{|S_0\eta/S_t|_{l^1(\mu)}\le n\}}\le0$ for all
$Z\in\Mc
_t^T(K^*\setminus\{0\})$ for all $n\ge1$. Moreover, $S_0g_n/S_T\in
L^\infty(l^1(\mu))$ for $n\ge1$. It then follows from (i) that the
sequence $(g_n)_{n\ge1}$ belongs to $\Xc_t^T$. Moreover, $g_n+\eta
S_T/S_t\in K_T$ for all $n\ge1$. Hence, $(g_n)_{n\ge1}$ $t$-Fatou
converges to $g$. Appealing to the $t$-Fatou closure property of
Theorem~\ref{thmSTFatouclosure} thus implies that $g\in\Xc_t^T$.

\item We then consider the case where $g\in L_{t,b}^0$ and is
such that $g^-:=((g^i)^-)_{i\ge1}$ satisfies $-g^-+\eta S_T/S_t\in
l^1_+(\mu)$ for some $\eta\in L^0(l^1_+(\mu), \Fc_t)$. We now
define the sequence $(g_n)_{n\ge1}$ by $g_n^i:=g^i\one_{\{g^i \le
n/(2^i\mu^i)\}}$ for $i\ge1$. It satisfies the requirement of (ii)
above and is $t$-Fatou convergent to $g$ since $g_n+ \eta S_T/S_t\ge
-g^-+\eta S_T/S_t\in l^1_+(\mu)\subset K_T$. Moreover, $\mathbb
{E}[Z_T\cdot g_n\mid\Fc_t]\le0$ for all $ Z\in\Mc_t^T(K^*\setminus\{
0\}
)$ since $g_n^i\le g^i$ for all $i\ge1$ and (\ref{eqpreuvecompliquee})
holds. By (ii), this implies that $g_n \in\Xc_{t}^T$
for all $n\ge1$. Since $\Xc_t^T$ is $t$-Fatou closed, by Theorem~\ref
{thmSTFatouclosure}, this implies that $g\in\Xc_t^T$.

\item We now turn to the case where $g\in L^0(l^1)$ and $g+\eta
S_T/S_t \in l^1_+$ for some $\eta\in L^0(l^1_+, \Fc_t)$. Let $\bar
\Mc_t^T$ denote the subset of elements $Z\in\Mc_t^T(K^*\setminus\{
0\})$ such that $Z^1_t=1$, fix $\eps>0$, and note that (\ref
{eqpreuvecompliquee}) implies that
%
%
\begin{equation}\label{eqpreuvecompliqueeeps}
\mathbb{E}[Z_T\cdot( g-\eps e_1S_T/S_t)\mid\Fc_t]\le-\eps
\qquad\mbox{for all } Z
\in\bar\Mc_t^T,
\end{equation}
since\vspace*{1pt} $Z\in\bar\Mc_t^T$ implies $\mathbb{E}[Z^1_TS^1_T/S^1_t\mid\Fc
_t]=Z^1_t=1$.
Let $g_{n}$ be defined by $g_{n}^i:=g^i\one_{\{g^i\ge0\mbox{ or }
i<n\}}$, $i\ge1$. Note that, for all $Z\in\bar\Mc_t^T$,
\begin{eqnarray*}
\mathbb{E}[Z_T\cdot(g_n-g)\mid\Fc_t]
&\le&
\mathbb{E}\biggl[\sum_{i\ge n} Z^i_T (g^i)^-\bigm|\Fc_t\biggr]\\
&\le&
\mathbb{E}\biggl[\sum_{i\ge n} Z^i_T\eta^i S^i_T/S^i_t\bigm|\Fc_t\biggr]
\\
&=&
\sum_{i\ge n} \eta^i Z^i_t,
\end{eqnarray*}
where the second inequality comes from the fact that $g+\eta S_T/S_t
\in l^1_+$ implies $(g^i)^-\le\eta^i S^i_T/S^i_t$ for all $i\ge1$.
Now observe that (\ref{eqbornesuplambda}) and (\ref{K*}) imply that
$0\le Z^i_t\le(1+c_t)$ for all $i\ge1$ and $Z \in\bar\Mc_t^T$, for
some $c_t\in L^0(\R,\Fc_t)$.
It then follows from the above inequalities, (\ref
{eqpreuvecompliqueeeps}) and the fact that $\eta\in l^1$ that
\[
\limsup_{n\to\infty} \esssup_{Z\in\bar\Mc_t^T}\mathbb
{E}[Z_T\cdot(g_n-\eps e_1S_T/S_t)\mid\Fc_t]\le-\eps.
\]
We can then find a sequence $(n_\eps)_{\eps>0}$ in $L^0(\N,\Fc_t)$
such that $n_\eps\to\infty$ a.s. as $\eps\to0$ and
\[
\mathbb{E}[Z_T\cdot(g_{n_\eps}-\eps e_1S_T/S_t)\mid\Fc_t]\le0
\qquad\mbox{for all } Z \in\bar\Mc_t^T.
\]
Moreover, $g_{n_\eps}-\eps e_1S_T/S_t$ satisfies the conditions of
(iii) above with $\eta_{n_\eps}:=(\eta^i\one_{i\le n_\eps})_{i\ge
1}+\eps e_1$ [recall (\ref{hypborneprix})] and therefore belongs to
$\Xc_t^T$ for all $\eps>0$. We conclude again by using the fact that
$\Xc_t^T$ is $t$-Fatou closed, by Theorem~\ref{thmSTFatouclosure},
and that $g_{n_\eps}+\eta S_T/S_t \in l^1_+\subset K_T$ for all
$\eps>0$.\qed
\end{longlist}
\noqed\end{pf*}

We conclude this section with the proof of Theorem~\ref{thmNA2equivaB}.
\begin{pf*}{Proof of Theorem~\ref{thmNA2equivaB}}
We follow the arguments of~\cite{DeKa10} which we adapt to our context.
Let us first fix an arbitrary $g\in(\xi{S_T/S_t} +\Xc_t^T)\cap
K_T$. In view of Lemma~\ref{rmsupmartproperty} applied with
$\eta=\xi$, one has $-Z_t\cdot\xi\le\mathbb{E}[Z_T\cdot g\mid\Fc
_t]\le0$
for all $Z \in\Mc_t^T(K^*\setminus\{0\})$. It then follows from \textbff
{B} that $\xi\in K_t$.

We now prove the converse assertion. Let us consider $\xi\in
L^0(l^1,\Fc_t)$ such that $Z_t\cdot\xi\ge0$ for all $Z\in\Mc
_t^T(K^*\setminus\{0\})$. We can then find $\alpha\in L^0(l^1_+,\Fc
_t)$ such that $-\xi+\alpha\in l_+^1$. By definition of $\Mc
_t^T(K^*\setminus\{0\})$, we have $0\le Z_t\cdot\xi=\mathbb
{E}[Z_T\cdot\xi S_T/S_t\mid\Fc_t]$ for all $Z \in\Mc
_t^T(K^*\setminus\{0\})$.
Moreover, $-\xi+\alpha\in l_+^1$ implies $- \xi S_T/S_t +\alpha
S_T/S_t\in l^1_+$, according to (\ref{hypborneprix}). It then
follows from Theorem~\ref{thmsuperhedging} applied to $g=-\xi
S_T/S_t$ that $-\xi S_T/S_t\in\Xc_t^T$. Hence, $0\in\xi S_T/S_t+\Xc
_t^T$, which by \textbff{NA2} implies that $\xi\in K_t$.
%
%
\end{pf*}

\subsection{A counter example}\label{subcontexemplefermeture}

In this section, we provide a counter example showing that Theorem \ref
{thmSTFatouclosure} can be false if Assumption~\ref{EffFric} is
replaced by a weaker one as in Remark~\ref{remcounterex}.

We consider a one-period model, $T=1$, in which $S_{0}=(1,1,\ldots)$,
$S^{1}_{1}=1$ and
\[
S^{i}_{1}:=U^{i} b^{i} + D^{i}(1-b^{i}),\qquad i\ge2,
\]
where $(b^{i})_{i\ge2}$ is a sequence of independent Bernoulli random
variables such that $\P[b^{i}=1]=1/2$, $U^{i}:=1+1/i$ and
$D^{i}:=1-1/i$, $i\ge2$. Note that each $S^{}$ is a martingale.

The transaction costs coefficients $\lambda^{ij}_{t}$ are defined by
$
\lambda^{1i}_{0}=\lambda^{i1}_{1}=\lambda^{ii}_{t}=0
$
for $i\ge1$ and $t=0,1$, and by
$
\lambda^{ij}_{0}=1/(i-1)
$
when $i\ge2$ and $i \neq j$,
$
\lambda^{ij}_{1}=1
$
when $j\ge2$ and $i \neq j$.

This market clearly satisfies (\ref{eqinegatriangulairelambda}), the
condition of Remark~\ref{remcounterex}
\[
\lambda^{ij}_t + \lambda^{ji}_t >0 \qquad\mbox{for all $t\in\T$ and
$i\neq j$,}
\]
and we shall show that it also satisfies the \textbff{NA2} Condition
\ref{CondNA2}. Indeed, by formula (\ref{K*}) one obtains that
\[
K^*_0=\{z \in l^\infty\dvtx z^1 \ge0, z^i \in z^1 [1-1/i,1], i
\geq2\}
\]
and
\[
K^*_1=\{z \in l^\infty\dvtx z^1 \ge0, z^i \in z^1 [1,2], i \geq
2\}.\vadjust{\goodbreak}
\]

In Condition~\ref{CondNA2}, the case $t=1$ is trivial. We next
consider the case $t=0$. Suppose that $\xi\in\Ac$, $\eta\in l^1$
and $(\eta+ \xi_0)S_1/S_0+\xi_1 \in L^0(K_1,\Fc_1)$. We must show
that $\eta\in K_0$. First note that $u:=(\eta+ \xi_0)S_1/S_0 \in
L^0(K_1,\Fc_1)$, by definition of $\Ac$, and thus satisfies $z \cdot
u \geq0$ for all $z \in K^*_1$, or equivalently, with $\alpha:=\eta+
\xi_0$,
\[
z \cdot u=\alpha^1 + \sum_{i \geq2} z^i \alpha^i (1+\epsilon^i/i)
\geq0\qquad \forall z^i \in[1,2], \epsilon^i \pm1.
\]
By choosing $z^i=1$ and $\epsilon^i=-1$ if $\alpha^i \ge0$, and,
$z^i=2$ and $\epsilon^i=+1$ if $\alpha^i < 0$ we obtain
\[
A:=\alpha^1 + \sum_{i \geq2}\bigl( \alpha^i_+ (1-1/i) -2\alpha
^i_- (1+1/i)\bigr) \geq0,
\]
where $a_+=\max\{0,a\}$ and $a_-=\max\{0,-a\}$.

With $B:=\alpha^1 +\sum_{i \geq2}( \alpha^i_+ (1-1/i)
-\alpha^i_- )$, we have $B\ge A$ and
\[
z \cdot\alpha= \alpha^1 + \sum_{i \geq2}\alpha^i z^i \ge B
\qquad\forall z \in K^*_0 \mbox{ with } z^1=1.
\]
This shows that $z \cdot\alpha\ge0$ for all $z \in K^*_0$, so
$\alpha\in K_0$. It then follows that $\eta\in K_0 -\xi_0 \subset
K_0$, which proves that \textbff{NA2} is satisfied.

We now show that $\Xc_{0}^{1}$ is not $0$-Fatou closed. To see this,
let us set
\[
h^{1}:= \sum_{i\ge2} y^{i} (2 b^{i}-1) \qquad\mbox{where } y^i=
i^{-(1+\epsilon)} \mbox{ for } i\ge2
\]
for some $\epsilon>0$.
We claim that, for each $n\ge1$, $g_{n}:=(h^{1}-n^{-1},0,0,\ldots)
\in\Xc_{0}^{1}$, while $g_{\infty}:=(h^{1},0,0,\ldots) \notin\Xc
_{0}^{1}$. Since $(g_{n})_{n}$ Fatou-converges to $g_{\infty}$, as a
uniformly bounded sequence in $L^{\infty}(l^{\infty})$ that converges
$\mbox{a.s.}$ pointwise, this shows that $\Xc_{0}^{1}$ is not Fatou-closed.

It remains to prove the above claims. We first show that $g_{n} \in\Xc
_{0}^{1}$. To see this, let us define
the sequence $\xi^{n}$ by
\begin{eqnarray*}
\xi_{0}^{n,i}&:=& \one_{2\le i\le I_{n}} i^{-\epsilon}- \one_{i=1}
\sum_{2\le j\le I_{n}} j^{-\epsilon}, \\
\xi_{1}^{n,i}&:=& - \one_{2\le i\le I_{n}} i^{-\epsilon} S^{i}_{1} +
\one_{i=1} \sum_{2\le j\le I_{n}} j^{-\epsilon} S^{j}_{1},\qquad i\ge1,
\end{eqnarray*}
where
\[
I_{n}:=\min\biggl\{k\ge2\dvtx\sum_{i\ge k} y^{i}(2b^{i}-1) \le n^{-1}\biggr\}.
\]
Note that $\xi^{n}\in\Ac$ by our choice of the structure of the
transaction costs. Moreover,
$
V^{\xi^{n}}_{1}=:(V^{n,1},0,0,\ldots)
$
with
\[
V^{n,1}= \sum_{2\le i\le I_{n}} i^{-\epsilon} (S^{i}_{1} -1)=2 \sum
_{2\le i\le I_{n}} y^{i}b^{i} - \sum_{2\le i\le I_{n}} y^{i}\ge
h^{1}-n^{-1},
\]
where we used the fact that $S^{i}_{1}-1=2b^{i}/i-1/i$.
This proves that $g_{n} \in\Xc_{0}^{1}$. We now show that $g_{\infty
} \notin\Xc_{0}^{1}$. Let $\tilde\Xc_{0}^{1}$ and $\tilde\Ac$ be
defined as $\Xc_{0}^{1}$ and $\Ac$ but for $\lambda= 0$. Clearly, $
\Xc_{0}^{1} \subset\tilde\Xc_{0}^{1}-L^{0}(\R^{\N}_{+})$, so that
it suffices to show that $g_{\infty} \notin\tilde\Xc
_{0}^{1}-L^{0}(\R^{\N}_{+})$.
Suppose that $g_{\infty} \in\tilde\Xc_{0}^{1}-L^{0}(\R^{\N
}_{+})$. Then one can find $\xi\in l^{1}$ and $c\in L^{0}(\R^{\N
}_{+})$ (recall that $S_{0}=1$) such that
\[
h^{1}= \sum_{ i\ge2} \xi^{i} (S^{i}_{1}-1 )-c^1.
\]
On the other hand
\begin{eqnarray*}
h^{1}&=&\sum_{i\ge2} y^{i} (2 b^{i}-1)\\
&=&\sum_{ i\ge2} \hat\xi^{i}
(S^{i}_{1}-1 )-\hat c \qquad\mbox{where } \hat c=0\mbox{ and
} \hat\xi^{i}:=i^{-\epsilon}\mbox{ for } i\ge2,
\end{eqnarray*}
where the above decomposition is unique in $\bigcup_{q<\infty
}l^{q}\times L^{0}(\R^{\N}_{+}) $, by independence of the Bernoulli
random variables $(b^{i})_{i\ge2}$. This is a contradiction since
$\hat\xi\notin l^{1}$, which proves that $g_{\infty} \notin\tilde
\Xc_{0}^{1}-L^{0}(\R^{\N}_{+})$.

\section{On the existence of many consistent price systems}\label{secmcps}

We split the proof of Theorem~\ref{thNA2equivMCPSeuivMSCPS} into
three parts. It follows from ideas introduced in~\cite{ras09} and~\cite{DeKa10} which we adapt to our context.
%
%
\begin{Theorem}\label{thmZtegaleta} Assume that \textup{\textbff{EF}} holds.
Then, \textup{\textbff{NA2}}${}\Rightarrow{}$\textup{\textbff{MCPS}}.
\end{Theorem}
\begin{pf}
We divide the proof into several points. In this proof, we use the
notation $F:=L^1(c_{0}(1/\mu))$ and $F':=L^\infty(l^1(\mu))$. From
now on, we fix $\eta\in L^0(\operatorname{int}{K^*_t})$ such that $\eta
S_t\in
L^1(l^\infty,\Fc_t)$. We set $G'=\R_+\eta$, which is the dual cone
of $G=\{y\dvtx y\in l^1, y\cdot x\ge0\ \forall x\in G'\}$. We also set
$\Theta:=(-L^0( G,\Fc_t)+ \Xc_t^T S_t/S_T)\cap F'$.\vspace*{8pt}

(1) We first show that $\Theta$ is $\sigma(F',F)$-closed. Let $B_1$
be the unit ball in $F'$. Arguing as in the proof of Theorem \ref
{thmfermeturefaible}, it suffices to show that, for any sequence
$(h_n)_{n\ge1} \subset\Theta\cap B_1$ that converges $\mbox{a.s.}$
to some
$h$, we have $h\in\Theta$. Let $(\zeta_n,V_n)_{n\ge1} \subset
-L^0(G,\Fc_t)\times\Xc_t^T$ be such that $\zeta_n +V_n S_t/S_T=h_n$
for all $n\ge1$. Since $h_n\in B_1$, we have $|h_n^i|\le1/\mu^i$ for
all $i\ge1$ and therefore $h_n+1/\mu\in l^1_+$ with $1/\mu\in
l^1_+$. It follows that $(\zeta_n+1/\mu)S_T/S_t+V_n=h_nS_T/S_t+(1/\mu
) S_T/S_t\in K_T$, which, by \textbff{NA2}, implies that $\zeta_n+1/\mu
\in K_t$. Since $\eta\in L^0(\operatorname{int}K^*_t,\Fc_t)$, we can find
$\eps\in L^0((0,1),\Fc_t)$ such that $\eta_n:=\eta- \eps(1_{\zeta
_n^i\ge0}-1_{\zeta_n^i < 0})_{i\ge1} \in K^*_t$ for all $n\ge1$. It
follows that $0\le\eta_n\cdot(\zeta_n+1/\mu)\le-\eps|\zeta
_n|_{l^1}+\eta\cdot\zeta_n+(\eta+\eps\one) \cdot1/\mu$. On the
other hand, we have $\eta\cdot\zeta_n \le0$ by definition of $G$
and $G'$. This shows that $(|\zeta_n|_{l^1})_{n\ge1} $ is $\mbox{a.s.}$
uniformly bounded. After possibly passing to ($\Fc_t$-measurable
random) subsequences (see the arguments used in the proof of Corollary
\ref{coroxinbornedansl1}), we can then assume that $(\zeta
_n)_{n\ge1}$ converges $\mbox{a.s.}$ in the product topology to some
$\zeta
\in L^0(l^1,\Fc_t)$. Moreover, we can find $(\alpha_n)_{n\ge1}
\subset L^0(l^1_+,\Fc_t)$ satisfying $\operatorname{ess}\sup_n |\alpha
_n|_{l^1}<\infty$ and such that $-\zeta_n+\alpha_n \in l^1_+$ for
all $n\ge1$. The identity $V_n =h_nS_T/S_t - \zeta_n S_T/S_t$ then
leads to $V_n +(1/\mu+\alpha_n) S_T/S_t\in K_T$ since $-\zeta
_n+\alpha_n \in l^1_+$ and $h_n+1/\mu\in l^1_+$. We conclude by
appealing to Theorem~\ref{thmSTFatouclosure}.

(2) We now show that $\Theta\cap L^0(\R^\N_+)=\{0\}$. Fix $(\zeta,
V) \in(-L^0( G,\Fc_t)\times\Xc_t^T)$ such that $\zeta+V S_t/S_T\in
\Theta\cap L^0(\R^\N_+)$. Then $\zeta S_T/S_t +V\in L^0(l^1_+)$, so
that $\zeta\in K_t$ by \textbff{NA2}. Since $\eta\in\operatorname{int}K^*_t$,
this implies that $\eta\cdot\zeta>0$ on $\{\zeta\ne0\}$. On the
other hand, the definition of $G$ and $G'$ leads to $\eta\cdot\zeta
\le0$. This shows that $\zeta=0$. An induction argument, based on
\textbff{NA2} and the fact that $K_s\cap(-K_s)=0$ for all $s\in\T$,
then implies that $V=0$.

(3) We can now complete the proof.
By the Hahn--Banach separation theorem, the fact that $\Theta$ is a
convex $\sigma(F',F)$-closed cone, that $\Theta\cap L^0(\R^\N_+)=\{
0\}$ and a standard exhaustion argument, we can find $Y\in F$ such that
$\mathbb{E}[Y\cdot h]\le0$ for all $h\in\Theta$, and $Y^i >0$ for
all $i
\ge1$. Defining the process $Z$ by $Z_s:=\mathbb{E}[YS_t\mid\Fc
_s]/S_s$ for
$t\le s\le T$, we obtain $Z^i >0$ for all $i \ge1$. Using the fact
that $-L^0( G,\Fc_t)\cap F' \subset\Theta$, we also obtain that
$Z_t\in G'$. From the fact that
$\Xc_t^T S_t/S_T \cap F' \subset\Theta$, we then deduce, as in the
proof of Proposition~\ref{propsepa1}, that $Z_s\in K^*_s$, for $t\le
s \le T$. Since $Z_t \in G'$, we can find a nonnegative $\Fc
_t$-measurable $\alpha$ such that $Z_t=\alpha\eta$.
Since $Z_t \neq0$, it follows that $\alpha>0$ a.s. Thus,
$(Z_s/\alpha)_{t\le s\le T}$ satisfies the required result.
\end{pf}
%
%
\begin{Lemma} \label{lmMCPSeuivMSCPS} Assume that \textup{\textbff{EF}} holds. Then,
\textup{\textbff{MCPS}}${}\Leftrightarrow{}$\textup{\textbff{MSCPS}}.
\end{Lemma}
\begin{pf}
As in~\cite{DeKa10}, we use a finite recursion from time $T$ to time
$0$ to prove that \textbff{MCPS}${}\Rightarrow{}$\textbff{MSCPS}. Let
\textbff{MSCPS}$(t)$ be the statement in \textbff{MSCPS} for $t\le T$
given. Suppose that \textbff{MCPS} is true. Then \textbff{MSCPS}$(T)$ is
trivially satisfied.

We now suppose that \textbff{MSCPS}$(s+1)$ is true for some $0 \leq s < T$.
Then, there exists an element $\tilde{X} \in\Mc_{s+1}^T(\operatorname
{int}K^*)$. Since $\tilde{X}_{s+1}S_{s+1} \in L^1(l^\infty)$, we can
define $\tilde X_s:=\mathbb{E}[\tilde{X}_{s+1}S_{s+1}\mid\Fc_s]/S_s$ and
$X_t:=\tilde X_t /(1+|\tilde X_s |_{l^\infty})$ for $s \leq t \leq T$.
Then $ 0< | X_s |_{l^\infty} <1$ and $X$ restricted to the interval
$(s, T]$ belongs to $\Mc_{s+1}^T(\operatorname{int}K^*)$.

Fix $\eta\in L^0(\operatorname{int}{K^*_s},\Fc_{s})$, let $d$ be its distance
to the border of $K^*_s$ and set $\alpha=(1 \wedge d)/2$. It follows
from formula (\ref{eqdistK*}) of Lemma~\ref{lmintK*} below that
$\alpha$ is $\Fc_s$-measurable. Since $ | X_s |_{\infty} <1$, we have
%
%
\begin{equation} \label{eq1MCPSeuivMSCPS}
\eta-\alpha X_s \in L^0(\operatorname{int}{K^*_s},\Fc_{s}).
\end{equation}
Let us now choose $\eta$ such that $\eta S_{s} \in L^1(l^\infty,\Fc
_s)$. Then $\eta S_{s} -\alpha X_s S_{s} \in L^1(l^\infty,\Fc_s)$,
and \textbff{MCPS} implies that there exists $Y \in\Mc
_{s}^T(K^*\setminus
\{0\})$ such that $Y_s = \eta-\alpha X_s$. In view of (\ref
{eq1MCPSeuivMSCPS}), $Y_s \in L^0(\operatorname{int}{K^*_s},\Fc_{s})$.

For $s \leq t \leq T$, define $Z_t=Y_t+\alpha X_t$. Then $Z_s =\eta\in
L^0(\operatorname{int}{K^*_s},\Fc_{s})$. Since, for $s+1 \leq t \leq
T$, $Y_t
\in L^0({K^*_t\setminus\{0\}},\Fc_{t})$ and $X_t \in L^0(\operatorname
{int}{K^*_t},\Fc_{t})$, and since $\alpha> 0$, it follows that $Z_t
\in L^0(\operatorname{int}{K^*_t},\Fc_{t})$ for such $t$. Hence $Z \in
\Mc
_{s}^T(\operatorname{int}K^*)$, so \textbff{MSCPS}$(s)$ is true.
\end{pf}
\begin{pf*}{Proof of Theorem~\ref{thNA2equivMCPSeuivMSCPS}}
In view of the above results, it remains to show that \textbff{MCPS}${}
\Rightarrow{}$\textbff{NA2}. Fix $\xi\in L^0(l^1,\Fc_t)\setminus
L^0(K_t,\Fc_t)$ such that $(\xi{S_T/S_t} +\Xc_t^T)\subset L^0(K_T)$.
Without loss of generality, we can assume that $\xi\in
L^\infty(l^1,\Fc_t)$, since otherwise we could replace $\xi$ by
$\xi/|\xi|_{l^1}$ and use the fact that $\Xc_t^T/|\xi|_{l^1}=\Xc_t^T$,
recall that $K$ is a cone valued process. It then follows from Lemma
\ref{rmsupmartproperty} that $0\ge- Z_t\cdot\xi$ for all $Z\in
\Mc_t^T(K^*\setminus\{0\})$. By the definition of \textbff{MCPS}, this
implies that $\eta\cdot\xi\ge0$ for all $\eta\in L^\infty(\operatorname
{int}{K^*_t},\Fc_{t})$. This shows that
$\xi\in K_t$.
\end{pf*}


\section{Elementary properties of $K$ and $K^*$}\label{seclast}

In this section, by a cone {is meant} a convex cone $C$ of vertex $0
\in C$, and \mbox{$(E,\|\cdot\|_E)$} denotes a Banach space with canonical
bilinear form $\sesq{\cdot}{\cdot}$. We recall that a cone $C$ in
$E$, is said to be normal (cf. Chapter V, Section 3.1 of \cite
{Schaefer}) if there exists $k \geq1$ such that
%
%
\begin{equation} \label{eqnormalcone}
\|x\|_E \leq k \|x+y\|_E\qquad \forall x,y \in C.
\end{equation}

The purpose of the first two results is to obtain that $K_t$ is normal
(a.s.) under \textbff{EF} and an explicit expression of the constant
$k$, used to establish measurability properties of the random cones
$K_t$ and $K^*_t$ and to establish bounds on order intervals defined by $K_t$.
%
%
\begin{Lemma}\label{lemknormcone}
Let $C$ be a cone in the Banach space $E$, and suppose that the dual cone
\[
C':=\{z\in E'\dvtx\sesq{z}{x}\ge0 \mbox{ for all } x \in C\}
\]
has an interior point $f_0$. Then $C$ is a normal cone and one can
choose $k=4 \|f_0\|_{E'}/d_{E'}(f_0,\partial C')$ in (\ref{eqnormalcone}).
\end{Lemma}
\begin{pf}
Let $d=d_{E'}(f_0,\partial C')$, and let $\bar B(a,r)$ denote the
closed ball in $E'$ of radius $r >0$ centered at $a$. We define
a norm $p$ in $E$ by
\[
p(x)=\sup\{|\sesq{f}{x}| \dvtx f \in\bar B(f_0,d)\},\qquad
x \in E.
\]
Substitution of $f=f_0 +d g$, $g \in\bar B(0,1)$ into this definition
and the fact that $d \leq\|f_0\|_{E'}$ give that $p(x) \leq
\|f_0\|_{E'} \|x\|_{E} + d \|x\|_{E} \leq2 \|f_0\|_{E'} \|x\|_{E}$. On
the other hand, we have
\[
\|x\|_{E}= \sup\{|\sesq{g}{x}| \dvtx g \in\bar B(0,1)\},
\]
which for $g=(f-f_0)/d\in\bar B(0,1)$ with $ f \in\bar B(f_0,d)$
similarly provides
\[
\|x\|_{E} \leq\sup\biggl\{\frac{1}{d}|\sesq{f}{x}|+\frac
{1}{d}|\sesq{f_0}{x}|\dvtx f \in\bar B(f_0,d)\biggr\} \leq\frac
{2}{d} p(x).\vadjust{\goodbreak}
\]
Hence $p(\cdot)$ and $\|\cdot\|_{E}$ are equivalent norms, since for
$x \in E$
\[
\frac{d}{2} \|x\|_{E} \leq p(x) \leq2 \|f_0\|_{E'} \|x\|_{E}.
\]

For $x,y \in C$, it follows directly from the fact that $ \bar
B(f_0,d)\subset C'$ and the definition of $p$ that $p(x+y) \geq p(x)$.
Then by the equivalence of the norms, for all $x,y \in C$,
\[
\|x\|_{E} \leq\frac{2}{d} p(x) \leq\frac{2}{d} p(x+y) \leq\frac
{4}{d}\|f_0\|_{E'} \|x+y\|_{E},
\]
which completes the proof by comparing with (\ref{eqnormalcone}).
\end{pf}
%
%
\begin{Lemma}\label{lembornel1siborneinfell2}
Let $C$ be a cone in the Banach space $E$, and suppose that $f_0$ is an
interior point of the dual cone $C'$. Then, there exists $ a >0$ such
that for all $y \in E$
\[
(C-y) \cap(y-C) \subset\bar B(0,a \sesq{f_0}{y}).
\]
%
Moreover (since $C$ is a normal cone), for any $k \ge1$ satisfying
(\ref{eqnormalcone}) and any $b \in(0,1)$, one can choose
\[
a=k/(b d_{E'}(f_0,\partial C')).
\]
\end{Lemma}
\begin{pf}
One observes that $x \in(C-y) \cap(y-C)$ if and only if $z_+ :=x+y
\in C$ and $z_- :=y-x \in C$. Since $C$ is normal according to Lemma
\ref{lemknormcone}, it follows that, for $\epsilon=\pm$,
\[
\|z_\epsilon\|_E \leq k \|z_+ + z_-\|_E = 2 k \|y \|_E.
\]
Then
\[
\|x\|_E = \tfrac{1}{2} \|z_+ - z_-\|_E \leq\tfrac{1}{2} (\|z_+\|_E +
\| z_-\|_E) \leq2 k \|y \|_E.
\]
Since $f_0$ is an interior point of $C'$, there exists $r>0$, such that
$f_0-rg \in C'$ for all $ g \in E'$ such that $\|g\|_{E'} \leq1$. For
$r>0$ sufficiently small, we thus have
\begin{eqnarray*}
\|y \|_E&=&\sup_{\|g \|_{E'} \leq1} | \sesq{g}{y}| =\sup_{\|g \|_{E'}
\leq1} \sesq{g}{y} = \sup_{g \in A_y}\sesq{g}{y} \\
&=&\frac{1}{r}\sup_{g \in A_y} (\sesq{f_0}{y}+ \sesq{rg-f_0}{y}) \leq
\frac{1}{r} \sesq{f_0}{y},
\end{eqnarray*}
where $A_y$ denotes the set of elements $g \in E'$ satisfying $\|g \|
_{E'} \leq1$ and \mbox{$\sesq{g}{y} \geq0$}, and
the last inequality follows from $f_0-rg \in C'$ while $y \in C$. This
shows that the inequality of the lemma is satisfied with $a=2k/r$. One
can choose $r=b d_{E'}(f_0,\partial C'))$ with $b \in(0,1)$, which
gives the stated choice of $a$.
\end{pf}

We now return to the particular case of $E=l^1$ and in the sequel of
this section, for ease of notation, we restrict to the case where
$\lambda$ is deterministic and constant in time. We therefore omit the
time index in $\lambda$, $K$ and $K^*$. We set $\Lambda:=(1+\lambda
)$ and use the notation
\[
\delta_u:=\inf_{i \neq j} (u^i\Lambda^{ij}- u^j)\qquad
\mbox{where } u \in l^\infty.
\]

%
\begin{Lemma} \label{lmintK*}
Assume that there exists some $c>0$ such that $\lambda^{ii}=0$ and $ 0
\leq\lambda^{ij} \leq c $ for all $i\ne j \geq1$.
Then, $u$ is an interior point of $K^*$ (in $l^{\infty}$) if and only
if $\delta_u >0$.

Suppose moreover that the interior of $K^*$ is nonempty. Then $u \in
\partial K^*$ if and only if $\delta_u=0$, $u \in l^\infty\setminus
K^*$ if and only if $\delta_u<0$ and the distance between a point $u
\in l^\infty$ and the border $\partial K^*$ is
%
%
\begin{equation} \label{eqdistK*}
d_{l^\infty}(u,\partial K^*)= \biggl|\inf_{i \neq j} \frac
{1}{1+\Lambda^{ij}}(u^i \Lambda^{ij}- u^j)\biggr|.
\end{equation}
\end{Lemma}
\begin{pf}
By definition, $u \in\operatorname{int}K^*$ if and only if $\exists
r>0$ such that $u+\bar{B}(0,r) \subset K^*$, where $\bar B(0,r)$
denotes the closed ball in $l^\infty$ centered at $0$ and with radius
$r$. Equivalently, {$z=u+|u|_{l^\infty}r'\epsilon$} satisfies (\ref
{K*}) for all $\epsilon\in\bar{B}(0,1)$, where {$r'=r/|u|_{l^\infty
}$} and $u \neq0$. For given $i \neq j$, choosing $\epsilon=-e_i+e_j$
leads to
%
%
\begin{equation}\label{eqinegapourint}
{r' |u|_{l^\infty}(1+\Lambda^{ij}) \leq u^i \Lambda^{ij} - u^j.}
\end{equation}
In particular, $\delta_u\ge r'|u|_{l^\infty}>0$ if $u \in
\operatorname{int}K^*$. Conversely, if $\delta_u>0$, then we can find
$r'>0$ such that (\ref{eqinegapourint}) holds. This implies that
\[
{u^j+|u|_{l^\infty}r'\le(u^i-|u|_{l^\infty}r')\Lambda^{ij},\qquad i,j\ge
1,}
\]
so that {$u+|u|_{l^\infty}r'\epsilon\in K^*$} for all $\epsilon\in
\bar{B}(0,1)$, that is, $u\in\operatorname{int}K^*$.

In the sequel of the proof, suppose that $\operatorname{int}K^*$ is
nonempty.
According to (\ref{K*}), $u \in K^*$ if and only if $\delta_u \geq0$,
and we have proved that $u\in\operatorname{int}K^*$ if and only if
$\delta_u
>0$. So it follows that $u \in l^\infty\setminus K^*$ if and only if
$\delta_u<0$ and that $u \in\partial K^*$ if and only if $\delta_u=0$.

It remains to prove (\ref{eqdistK*}). Let $d$ denote the right-hand
side of (\ref{eqdistK*}). Suppose first that $\delta_u >0$. For all
$\delta>0$ we can choose $i\ne j$ such that $\frac{1}{1+\Lambda
^{ij}}(u^i \Lambda^{ij}- u^j)<d+\delta$. Then, $\delta
_{u+(d+\delta) (-e_i+e_j)} < 0$, so $u+(d+\delta) (-e_i+e_j) \notin
K^*$. This shows that $d_{l^\infty}(u,\partial K^*)\le d$. Conversely,
for all $\epsilon\in\bar B(0,1)$ $\delta_{u+d \epsilon} \geq0$, so
$u+d \epsilon\in K^*$. Hence, $d\le d_{l^\infty}(u,\partial K^*)$
which proves (\ref{eqdistK*}), when $\delta_u >0$. Proceeding
similarly, we obtain for the case $\delta_u <0$ that $\delta_{u+d
\epsilon} \leq0$ for all $\epsilon\in\bar B(0,1)$, and that for all
$\delta>0$ there exists $i\neq j$ such that $\delta_{u+(d+\delta)
(e_i-e_j)} > 0$. To complete the proof we note that (\ref{eqdistK*})
gives $d_{l^\infty}(u,\break\partial K^*)=0$, when $\delta_u =0$.
\end{pf}
%
%
\begin{Proposition} \label{propEFNormCone}
Assume that there exists some $c>0$ such that $\lambda^{ii}=0$ and $ 0
\leq\lambda^{ij} \leq c $ for all $i\ne j \geq1$. Then, the
following assertions:

{\renewcommand\thelonglist{(\arabic{longlist})}
\renewcommand\labellonglist{\thelonglist}
\begin{longlist}
\item\label{propEFNormCone1} $\exists\eps> 0$ such that
$\lambda^{ij} \geq\eps$ $\forall i\neq j;$
\item\label{propEFNormCone2} $\one$ is an interior point of $K^*$;\vadjust{\goodbreak}
\item\label{propEFNormCone3} $K$ is a normal cone;
\item\label{propEFNormCone4} $K^*$ has the generating
property, that is, $l^\infty= K^*-K^*$;
\item\label{propEFNormCone5} $\exists\eps> 0$ such that
$\lambda^{ij} + \lambda^{ji} \geq\eps$ $\forall i\neq j$,
\end{longlist}}

\noindent satisfy:~\ref{propEFNormCone1} $\Leftrightarrow$ \ref
{propEFNormCone2} $\Rightarrow$~\ref{propEFNormCone3} $\Leftrightarrow
$~\ref{propEFNormCone4} $\Rightarrow$~\ref{propEFNormCone5}.
\end{Proposition}
\begin{pf}
The equivalence of~\ref{propEFNormCone1} and~\ref{propEFNormCone2} is
a direct consequence of Lem\-ma~\ref{lmintK*}. The
equivalence between~\ref{propEFNormCone3} and~\ref{propEFNormCone4}
is standard; cf. Chapter V, Section 3.5 of~\cite{Schaefer}.

In the rest of the proof, we shall use the following notation:
\[
f_{ij}:=\Lambda^{ij} e_i-e_j \qquad\mbox{for } i\ne j\ge1,
\qquad x:=\sum
_{i \neq j} a^{ij} f_{ij} \quad\mbox{and}\quad y:=\sum_{i \neq j} b^{ij}
f_{ij},
\]
where $a, b \in\M_{f,+}$ will be given by the context.

We now prove that~\ref{propEFNormCone1} implies~\ref{propEFNormCone3}.
Since $x=\sum_{i \neq j} (\Lambda^{ij} a^{ij}- a^{ji})e_i$ and
$|f_{ij}|_{l^1} = \Lambda^{ij} +1$, we have
\[
\sum_{i \neq j} (\Lambda^{ij}-1) a^{ij} \leq|x|_{l^1} \leq\sum_{i
\neq j} (\Lambda^{ij}+1) a^{ij} \leq(2+c) \sum_{i \neq j} a^{ij}.
\]
Then, according to the above inequality,
\[
\eps\sum_{i \neq j} a^{ij} \leq|x|_{l^1} \leq(2+c) \sum_{i \neq j} a^{ij}.
\]
Similarly,
\[
\eps\sum_{i \neq j}( a^{ij}+b^{ij}) \leq|x+y|_{l^1}.
\]
Combining the above inequalities leads to
\[
|x|_{l^1} \leq(2+c) \sum_{i \neq j} a^{ij} \leq(2+c) \sum_{i \neq
j} (a^{ij} +b^{ij})\leq\frac{2+c}{\eps} |x+y|_{l^1}.
\]
It then follows that
\[
|x|_{l^1} \leq\frac{2+c}{\eps} |x+y|_{l^1}
\]
for all $x,y \in K$, which proves that $K$ is normal.

It remains to prove that~\ref{propEFNormCone3} implies \ref
{propEFNormCone5}. Let us assume that the condition \ref
{propEFNormCone3} is satisfied. Let $x$ and $y$ be defined as above with
$a,b\in\M_{f,+}$ such that $b^{ij}=a^{ji}$ for all $i,j\ge1$, and
set $d^{ij}:=a^{ij}+b^{ij}=a^{ij}+a^{ji}$, so that $d^{ij}=d^{ji}$, and
$x+y= \sum_{i \neq j} d^{ij} (\Lambda^{ij} -1)e_i$. Then,
\[
|x+y|_{l^1}= \sum_{i \neq j} d^{ij} (\Lambda^{ij} -1)=\frac
{1}{2}\sum_{i \neq j} d^{ij} (\lambda^{ij} + \lambda^{ji})=\sum_{i
\neq j} a^{ij} (\lambda^{ij} + \lambda^{ji}).
\]
Since $K$ is normal, there is $k\geq1$, independent on $x$ and $y$,
such that $|x|_{l^1} \leq k |x+y|_{l^1}$, which, combined with the
previous inequality, implies
\[
|x|_{l^1} \leq k \sum_{i \neq j} a^{ij} (\lambda^{ij} + \lambda^{ji}).
\]
Considering the case where $x=f_{mn}$ for some $m\ne n$, then leads to
$2+\lambda^{mn} \leq k (\lambda^{mn} + \lambda^{nm})$. It follows
that $ \lambda^{mn} + \lambda^{nm} \geq2/k$, which, by the
arbitrariness of $(m,n)$, proves that~\ref{propEFNormCone5} is satisfied.
\end{pf}
%
%
\begin{Remark}\label{remlambdaij+ji}
Assertion~\ref{propEFNormCone5} of Proposition \ref
{propEFNormCone} does not imply that $K$ is normal [assertion \ref
{propEFNormCone3}], or equivalently that $K^*$ has the generating property~\ref{propEFNormCone4}. 
Since $\operatorname{int}K^*\ne\varnothing$ implies
that $K^*$ has the generating property, this shows that \ref
{propEFNormCone5} does not imply that $\operatorname{int}K^*\ne
\varnothing$. An
example is given by the case where $\lambda^{ij}=1$ for $i<j$ and
$\lambda^{ij}=0$ for $i\geq j$.

Indeed, assume that $\lambda$ satisfies the above condition, let $x
\in l^{\infty}$ be defined by $x=(1,0,1,0, \ldots)$ and suppose that
it can be written as $x=y_1-y_2$, for some $y_1,y_2 \in K^*$.
First note that the definition of $\lambda$ implies that
%
%
\begin{equation}\label{eqcontraproprgene}
\mbox{$0 \leq y^j \leq y^i \leq2 y^j$}\qquad\mbox{for $j<i$ whenever $y \in K^*$.}
\end{equation}
In view of the left-hand side of (\ref{eqcontraproprgene}) and the
identity $x=y_1-y_2$, we should then have $y_1^{2n-1}=a^{2n-1} + n$,
$y_1^{2n}=a^{2n} + n$, $y_2^{2n-1}=a^{2n-1} + n-1$ and $y_2^{2n}=a^{2n}
+ n$ for $n \geq1$, where $(a^n)_{n\ge1}$ is an increasing
nonnegative sequence. On the other hand, the right-hand side of (\ref
{eqcontraproprgene}) implies that $0 \leq y^i\leq2 y^1$ for $i>1$.
This leads to a contradiction, therefore showing that $x\notin
K^*-K^*$, that is, that the generating property is not satisfied.
\end{Remark}
%


\section{Concluding remarks}\label{secconcludingremars}

Our main results could be obtained in a more abstract setting as
described below.

Let us consider the situation where $(K_{t})_{t\in\T}$ is just
assumed to be a family of random cones, {together} with the following
properties, for $\P$-a.e. $\omega\in\Omega$ and all $t\in\T$:

\begin{longlist}[(iii)]
\item[(i)] $K_t(\omega)$ is a closed convex cone in $l^1$ of vertex
$0$ satisfying $l_+^1 \subset K_t(\omega)$. {The dual cone
$K^*_t(\omega)$ has an interior point $\theta_t(\omega)$ such that
$\theta_t \in L^0(l^\infty,\Fc_t)$.}
\item[(ii)] $d_{l^{\infty}}(\theta_{t},\partial K^{*}_{t})\in
L^{0}((0,\infty),\Fc_{t})$.
\item[(iii)] 
There exists a family $\Ec_{t}\subset L^{\infty}(K_{t}\cap c_{f})$
such that $K^{*}_{t}(\omega)=\{z\in l^{\infty}\dvtx\break z\cdot\zeta
_{t}(\omega)\ge0$ for all $\zeta_{t}\in\Ec_{t}\}$.
%
\item[(iv)] 
There exist{s} a constant $C$, independent o{f} $\omega$, such that $z
\in K^{*}_{t}(\omega)\Rightarrow|z^{i}|\le C(1+|z^{1}|)$ for
all $i\ge1$.
\end{longlist}

The proofs of Theorems~\ref{thmSTFatouclosure} and \ref
{thmfermeturefaible} only appeal to (i) and (ii) above.
The proof of Proposition~\ref{eqinclusiontildeXc} is adapted under
(iii) by replacing the simple elements $-\alpha e_i S^i_0/S^i_t\chi_{\{
S^i_t\ge\eps\}} $ and $\alpha(e_j-(1+\lambda^{ij}_t) e_i)
S_0/S_t\chi_{\{S^j_t\wedge S^i_t\ge\eps\}}$ by 
$-\alpha\zeta_{t}S_{0}/S_{t}\chi_{E_{\zeta_{t}}}$ where $E_{\zeta
_{t}}:=\{S^{j}_{t}\ge\eps$, for all $j\ge1$ such that $\zeta
_{t}^{j}\ne0\}$
for $\zeta_{t}\in\Ec_{t}$. Hence, Proposition~\ref{eqinclusiontildeXc}
remains true under 
(i), (ii) and (iii).
If we now add (iv) as an assumption, one can repeat the arguments of
the proof of Corollary~\ref{coroMtTnonvide}. No other modification
is then required to prove Theorem~\ref{thmsuperhedging}. Theorems \ref
{thmNA2equivaB} and~\ref{thNA2equivMCPSeuivMSCPS}
similarly hold under (i)--(iv).

In the case where $\Ec_{t}$ is countable, $\Ec_{t}=\{\zeta_{it}$,
$i\ge1\}$, the properties (i), (ii) and (iii) are not independent. An
adapted version of Lemma~\ref{lmintK*} is indeed true with minor
changes: $d_{l^\infty}(u,\partial K^*)= |{\inf_{i \geq1} \frac
{1}{|\zeta_i|_{l^{1}}} (u \cdot\zeta_i)}|$. It follows that
(i) and (iii) implies (ii) in this case.

As explained in the \hyperref[secintroduction]{Introduction}, we have
considered here a model in which financial strategies are described by
amounts of money as opposed to number of units.
The main reason is that, in the latter setting, our assumption \textbff{EF}
would impose a strong nondegeneracy condition on the bid ask matrices
$(\pi_{t}^{ij})_{ij}$. Note also that the linear function $x \mapsto
Sx$ does not define an isomorphism of ``nice'' TVS, so that there is no
such natural way to pass from a model in amounts to a model in quantities.
Obviously, from the pure mathematical point of view, one can always
consider an abstract family of cones, as described above, and set
$S\equiv1$, so as to recover a general model for strategies {labeled}
in terms of units.


%

\printaddresses


\begin{thebibliography}{24}

\bibitem{BMKR97}
%
\begin{barticle}[auto:STB|2012/04/12|05:18:16]
\bauthor{\bsnm{Bjork},~\bfnm{T.}\binits{T.}},
\bauthor{\bsnm{Di~Masi},~\bfnm{G.}\binits{G.}},
\bauthor{\bsnm{Kabanov},~\bfnm{Y.}\binits{Y.}} \AND
\bauthor{\bsnm{Runggaldier},~\bfnm{W.}\binits{W.}}
(\byear{1997}).
\btitle{Towards a general theory of bond markets}.
\bjournal{Finance Stoch.}
\bvolume{1}
\bpages{141--174}.
\bptok{imsref}%
\end{barticle}
%
\endbibitem

\bibitem{BoCh09}
%
\begin{barticle}[mr]
\bauthor{\bsnm{Bouchard},~\bfnm{Bruno}\binits{B.}} \AND
\bauthor{\bsnm{Chassagneux},~\bfnm{Jean-Fran{\c{c}}ois}\binits{J.-F.}}
(\byear{2009}).
\btitle{Representation of continuous linear forms on the set of ladlag
processes and the hedging of {A}merican claims under proportional costs}.
\bjournal{Electron. J. Probab.}
\bvolume{14}
\bpages{612--632}.
\bid{issn={1083-6489}, mr={2486816}}
\bptok{imsref}%
\end{barticle}
%
\endbibitem

\bibitem{campischach}
%
\begin{barticle}[mr]
\bauthor{\bsnm{Campi},~\bfnm{Luciano}\binits{L.}} \AND
\bauthor{\bsnm{Schachermayer},~\bfnm{Walter}\binits{W.}}
(\byear{2006}).
\btitle{A super-replication theorem in {K}abanov's model of transaction costs}.
\bjournal{Finance Stoch.}
\bvolume{10}
\bpages{579--596}.
\bid{doi={10.1007/s00780-006-0022-4}, issn={0949-2984}, mr={2276320}}
\bptok{imsref}%
\end{barticle}
%
\endbibitem

\bibitem{Carmona-Tehr}
%
\begin{barticle}[mr]
\bauthor{\bsnm{Carmona},~\bfnm{Rene}\binits{R.}} \AND
\bauthor{\bsnm{Tehranchi},~\bfnm{Michael}\binits{M.}}
(\byear{2004}).
\btitle{A characterization of hedging portfolios for interest rate contingent
claims}.
\bjournal{Ann. Appl. Probab.}
\bvolume{14}
\bpages{1267--1294}.
\bid{doi={10.1214/105051604000000297}, issn={1050-5164}, mr={2071423}}
\bptok{imsref}%
\end{barticle}
%
\endbibitem

\bibitem{DP05b}
%
\begin{barticle}[mr]
\bauthor{\bsnm{De~Donno},~\bfnm{M.}\binits{M.}} \AND
\bauthor{\bsnm{Pratelli},~\bfnm{M.}\binits{M.}}
(\byear{2005}).
\btitle{A theory of stochastic integration for bond markets}.
\bjournal{Ann. Appl. Probab.}
\bvolume{15}
\bpages{2773--2791}.
\bid{doi={10.1214/105051605000000548}, issn={1050-5164}, mr={2187311}}
\bptok{imsref}%
\end{barticle}
%
\endbibitem

\bibitem{DeDeKa}
%
\begin{barticle}[mr]
\bauthor{\bsnm{De~Valli{\`e}re},~\bfnm{D.}\binits{D.}},
\bauthor{\bsnm{Denis},~\bfnm{E.}\binits{E.}} \AND
\bauthor{\bsnm{Kabanov},~\bfnm{Y.}\binits{Y.}}
(\byear{2009}).
\btitle{Hedging of {A}merican options under transaction costs}.
\bjournal{Finance Stoch.}
\bvolume{13}
\bpages{105--119}.
\bid{doi={10.1007/s00780-008-0076-6}, issn={0949-2984}, mr={2465488}}
\bptok{imsref}%
\end{barticle}
%
\endbibitem

\bibitem{DeKa10}
%
\begin{barticle}[auto:STB|2012/04/12|05:18:16]
\bauthor{\bsnm{Denis},~\bfnm{E.}\binits{E.}} \AND
\bauthor{\bsnm{Kabanov},~\bfnm{Y.}\binits{Y.}}
(\byear{2012}).
\btitle{Consistent price systems and arbitrage opportunities of the
second kind in models with transaction costs}.
\bjournal{Finance Stoch.}
\bvolume{16}
\bpages{135--154}.
\bid{mr={2872651}}
\bptok{imsref}%
\end{barticle}
%
\endbibitem

\bibitem{ET}
%
\begin{barticle}[mr]
\bauthor{\bsnm{Ekeland},~\bfnm{Ivar}\binits{I.}} \AND
\bauthor{\bsnm{Taflin},~\bfnm{Erik}\binits{E.}}
(\byear{2005}).
\btitle{A theory of bond portfolios}.
\bjournal{Ann. Appl. Probab.}
\bvolume{15}
\bpages{1260--1305}
\bnote{(cf. arXiv:\arxivurl{math/0301278v3} [math.OC]).}
\bid{doi={10.1214/105051605000000160}, issn={1050-5164}, mr={2134104}}
\bptok{imsref}%
\end{barticle}
%
\endbibitem

\bibitem{GrepaKaba}
%
\begin{bmisc}[auto:STB|2012/04/12|05:18:16]
\bauthor{\bsnm{Gr{\'e}pat},~\bfnm{J.}\binits{J.}} \AND
\bauthor{\bsnm{Kabanov},~\bfnm{Y.}\binits{Y.}}
(\byear{2010}).
\bhowpublished{Small transaction costs, absence of arbitrage and consistent
price systems. Preprint.}
\bptok{imsref}%
\end{bmisc}
%
\endbibitem

\bibitem{GuRaSc08}
%
\begin{barticle}[mr]
\bauthor{\bsnm{Guasoni},~\bfnm{Paolo}\binits{P.}},
\bauthor{\bsnm{R{\'a}sonyi},~\bfnm{Mikl{\'o}s}\binits{M.}} \AND
\bauthor{\bsnm{Schachermayer},~\bfnm{Walter}\binits{W.}}
(\byear{2010}).
\btitle{The fundamental theorem of asset pricing for continuous processes under
small transaction costs}.
\bjournal{Annals of Finance}
\bvolume{6}
\bpages{157--191}.
\bptok{imsref}%
\end{barticle}
%
\endbibitem

\bibitem{JKequi}
%
\begin{barticle}[mr]
\bauthor{\bsnm{Jouini},~\bfnm{Elyes}\binits{E.}} \AND
\bauthor{\bsnm{Kallal},~\bfnm{H{\'e}di}\binits{H.}}
(\byear{1995}).
\btitle{Martingales and arbitrage in securities markets with transaction
costs}.
\bjournal{J. Econom. Theory}
\bvolume{66}
\bpages{178--197}.
\bid{doi={10.1006/jeth.1995.1037}, issn={0022-0531}, mr={1338025}}
\bptok{imsref}%
\end{barticle}
%
\endbibitem

\bibitem{KSR01}
%
\begin{barticle}[mr]
\bauthor{\bsnm{Kabanov},~\bfnm{Yuri}\binits{Y.}},
\bauthor{\bsnm{R{\'a}sonyi},~\bfnm{Mikl{\'o}s}\binits{M.}} \AND
\bauthor{\bsnm{Stricker},~\bfnm{Christophe}\binits{C.}}
(\byear{2002}).
\btitle{No-arbitrage criteria for financial markets with efficient friction}.
\bjournal{Finance Stoch.}
\bvolume{6}
\bpages{371--382}.
\bid{doi={10.1007/s007800100062}, issn={0949-2984}, mr={1914317}}
\bptok{imsref}%
\end{barticle}
%
\endbibitem

\bibitem{KSR04}
%
\begin{barticle}[mr]
\bauthor{\bsnm{Kabanov},~\bfnm{Yuri}\binits{Y.}},
\bauthor{\bsnm{R{\'a}sonyi},~\bfnm{Mikl{\'o}s}\binits{M.}} \AND
\bauthor{\bsnm{Stricker},~\bfnm{Christophe}\binits{C.}}
(\byear{2003}).
\btitle{On the closedness of sums of convex cones in {$L\sp0$} and the robust
no-arbitrage property}.
\bjournal{Finance Stoch.}
\bvolume{7}
\bpages{403--411}.
\bid{doi={10.1007/s007800200089}, issn={0949-2984}, mr={1994916}}
\bptok{imsref}%
\end{barticle}
%
\endbibitem

\bibitem{Kab-Saf}
%
\begin{bbook}[mr]
\bauthor{\bsnm{Kabanov},~\bfnm{Yuri}\binits{Y.}} \AND
\bauthor{\bsnm{Safarian},~\bfnm{Mher}\binits{M.}}
(\byear{2009}).
\btitle{Markets with Transaction Costs, Mathematical Theory}.
\bpublisher{Springer}, \baddress{Berlin}.
\bid{doi={10.1007/978-3-540-68121-2}, mr={2589621}}
\bptok{imsref}%
\end{bbook}
%
\endbibitem

\bibitem{teachersnote}
%
\begin{bincollection}[mr]
\bauthor{\bsnm{Kabanov},~\bfnm{Yuri}\binits{Y.}} \AND
\bauthor{\bsnm{Stricker},~\bfnm{Christophe}\binits{C.}}
(\byear{2001}).
\btitle{A teachers' note on no-arbitrage criteria}.
In \bbooktitle{S\'eminaire de {P}robabilit\'es, {XXXV}}.
\bseries{Lecture Notes in Math.}
\bvolume{1755}
\bpages{149--152}.
\bpublisher{Springer}, \baddress{Berlin}.
\bid{mr={1837282}}
\bptok{imsref}%
\end{bincollection}
%
\endbibitem


\bibitem{Pham2003}
%
\begin{barticle}[mr]
\bauthor{\bsnm{Pham},~\bfnm{Huy{\^e}n}\binits{H.}}
(\byear{2003}).
\btitle{A predictable decomposition in an infinite assets model with jumps.
{A}pplication to hedging and optimal investment}.
\bjournal{Stoch. Stoch. Rep.}
\bvolume{75}
\bpages{343--368}.
\bid{doi={10.1080/104511203100001621237}, issn={1045-1129}, mr={2017783}}
\bptok{imsref}%
\end{barticle}
%
\endbibitem

\bibitem{ras09}
%
\begin{bincollection}[mr]
\bauthor{\bsnm{R{\'a}sonyi},~\bfnm{Mikl{\'o}s}\binits{M.}}
(\byear{2009}).
\btitle{Arbitrage under transaction costs revisited}.
In \bbooktitle{Optimality and Risk---Modern Trends in Mathematical Finance}
(\beditor{F. Delbaen}, \beditor{M. R\'{a}sonyi} and \beditor{C.
Stricker}, eds.)
\bpages{211--225}.
\bpublisher{Springer}, \baddress{Berlin}.
\bid{doi={10.1007/978-3-642-02608-9_11}, mr={2648605}}
\bptok{imsref}%
\end{bincollection}
%
\endbibitem

\bibitem{schach}
%
\begin{barticle}[mr]
\bauthor{\bsnm{Schachermayer},~\bfnm{Walter}\binits{W.}}
(\byear{2004}).
\btitle{The fundamental theorem of asset pricing under proportional transaction
costs in finite discrete time}.
\bjournal{Math. Finance}
\bvolume{14}
\bpages{19--48}.
\bid{doi={10.1111/j.0960-1627.2004.00180.x}, issn={0960-1627}, mr={2030834}}
\bptok{imsref}%
\end{barticle}
%
\endbibitem

\bibitem{Schaefer}
%
\begin{bbook}[mr]
\bauthor{\bsnm{Schaefer},~\bfnm{Helmut~H.}\binits{H.~H.}}
(\byear{1999}).
\btitle{Topological Vector Spaces}, \bedition{2nd} ed.
\bpublisher{Springer}, \baddress{New York}.
\bptnote{check year}%
\bptok{imsref}%
\end{bbook}
%
\endbibitem

\bibitem{ETBondCompleteness}
%
\begin{barticle}[mr]
\bauthor{\bsnm{Taflin},~\bfnm{Erik}\binits{E.}}
(\byear{2005}).
\btitle{Bond market completeness and attainable contingent claims}.
\bjournal{Finance Stoch.}
\bvolume{9}
\bpages{429--452}
\bnote{(cf. arXiv:\arxivurl{math/0402364v2} [math.OC]).}
\bid{doi={10.1007/s00780-005-0156-9}, issn={0949-2984}, mr={2211717}}
\bptok{imsref}%
\end{barticle}
%
\endbibitem

\bibitem{T09}
%
\begin{barticle}[mr]
\bauthor{\bsnm{Taflin},~\bfnm{Erik}\binits{E.}}
(\byear{2011}).
\btitle{Generalized integrands and bond portfolios: Pitfalls and counter
examples}.
\bjournal{Ann. Appl. Probab.}
\bvolume{21}
\bpages{266--282}
\bnote{(cf. arXiv:\arxivurl{0909.2341v2} [math.PR]).}
\bid{doi={10.1214/10-AAP694}, issn={1050-5164}, mr={2759202}}
\bptok{imsref}%
\end{barticle}
%
\endbibitem

\bibitem{Yosida}
%
\begin{bbook}[mr]
\bauthor{\bsnm{Yosida},~\bfnm{K{\^o}saku}\binits{K.}}
(\byear{1974}).
\btitle{Functional Analysis},
\bedition{4th} ed.
\bseries{Die Grundlehren der Mathematischen Wissenschaften}
\bvolume{123}.
\bpublisher{Springer}, \baddress{New York}.
\bid{mr={0350358}}
\bptok{imsref}%
\end{bbook}
%
\endbibitem

\end{thebibliography}
\end{document}